\documentclass[a4paper,12pt]{article}

\usepackage[T2A]{fontenc}
\usepackage[utf8]{inputenc}

\usepackage{amsmath,amssymb}
\usepackage{graphicx}

\numberwithin{equation}{section}

\usepackage[english]{babel}
\usepackage{amsfonts}
\usepackage{epsfig}
\usepackage{booktabs}
\usepackage{bbold}
\usepackage{graphicx}
\DeclareGraphicsExtensions{.eps,.ps,.jpg,.bmp}

\newcommand{\nnr}{\nonumber\\}
\newcommand{\ex}{{\rm e}}
\newtheorem{Theorem}{Proposition}

\newtheorem{Definition}{Conjecture}

\newcommand{\CC}{{\mathbb C}}

\newcommand{\fr}{\frac}
\newcommand{\pt}{\partial}

\newcommand{\eps}{\epsilon}

\title{}
\author{}
\date{}
\begin{document}
\begin{titlepage}

\hfill\parbox{40mm}

\vspace{30mm}

\begin{center}
{\large \bf Solutions of Kapustin-Witten equations for $ADE$-type  groups  }

\vspace{17mm}

\textrm{Zhi Sheng Liu$^1$ and Bao Shou$^2$}
\vspace{8mm}

\textit{$^2$Center  of Mathematical  Sciences\\
Zhejiang University \\
Hangzhou, 310027, China}\\

\textit{$^1$Institute  of Theoretical  Physics\\
Chinese Academy  of Sciences\\
Beijing, 100190, China}\\

\texttt{zsliu@itp.ac.cn, bsoul@zju.edu.cn}

\vspace{3.5cm}

{\bf Abstract}\end{center}
Kapustin-Witten (\textbf{KW}) equations are encountered in the localization of the topological ${\mathcal N}=4$ SYM theory.
 Mikhaylov has constructed model solutions of \textbf{KW} equations for the boundary 't~Hooft  operators on a half space. Direct  proof of the solutions boils down to check a boundary condition.  There are two computational   difficulties in  explicitly constructing the solutions for  higher rank Lie algebra. The first one is related to the commutation  of generators of Lie algebra. We derive  an identity which effectively reduces this computational difficulty.  The second  one involves the number of ways from the highest  weights to other weights in the fundamental representation.   For $ADE$-type gauge groups, we find an amazing formula which can be used to rewrite the   solutions of \textbf{KW} equations. This new formula of solutions   bypass above two computational difficulties. We also discuss this formula for all minuscule representations and none simple lattice Lie algebras.
\end{titlepage}
\section{Introduction}
The maximally supersymmetric Yang-Mills theory in four dimensions can be twisted in three ways to obtain topological field theories. One of the twists called the GL twist\cite{Langlands} appears to be relevant for the geometric Langlands program. It can be  applied to  the description of the Khovanov homology of knots \cite{ComplexCS,5knots, GaiottoWitten}.  The Chern-Simons theory is effectively induced on the boundary of a four-dimensional manifold. The supersymmetry conditions lead to the generalized Bogomolny equations \cite{Langlands} which is called Kapustin-Witten (\textbf{KW}) equations now.

As described in \cite{5knots}, on a half space $V$ of the form $V=\mathbb{R}^3\times \mathbb{R}_+$, the \textbf{KW} equations  are
\begin{eqnarray}
F-\phi\wedge\phi+\ast{ d}_A\phi =0=d_A\ast\phi\,,\label{bogomol1}
\end{eqnarray}
where $d_A$ is the covariant exterior derivative associated with a connection $A$, and $\phi$ is  one-form valued in the adjoint of the gauge group $G$. Different reductions of the \textbf{KW} equations lead to other well known equations {\it e.g.}, Nahm's equations, Bogomolny equations or Hitchin equations.  Through electric-magnetic duality, the natural Chern-Simons observables correspond to the boundary 't~Hooft or surface operators in four dimensional gauge theory. These operators are defined by prescribing the singular behavior of the fields  as the supersymmetry  boundary conditions in the model.

These model solutions  with 't~Hooft operator as boundary conditions were first discussed  in \cite{5knots} for $SU(2)$ gauge group. The boundary conditions and  solutions  were studied further in \cite{Henningson2}. For higher rank groups,  solutions  were constructed for special values of the magnetic weight in \cite{Henningson}.  For any simple compact gauge group,  after reducing to a Toda system,  \cite{vm} V.Mikhaylov conjectured a formula of the model solutions  for the boundary 't~Hooft operator with general magnetic weight.  Model solutions  for the $SU(n)$ groups were also obtained in \cite{vm} for the boundary surface operator. For other related  work on these equations, see\cite{uhlenbeck}\cite{M-Witten}\cite{He}\cite{Yuuji}\cite{Teng}.

Proof of the conjecture of the solutions requires to check a boundary condition. This has been completed for $SU(n)$ group in \cite{vm}. In order to  check the boundary condition, we need   construct the  solutions explicitly. Unfortunately, there are two 'NP'-like computational difficulties with the increasing rank of Lie algebra. One difficulty is related to  the commutation  of generators  of Lie algebra.  Another difficulty involves the number of  paths from the highest weight to an arbitrarily weight in  the fundament representation. The purpose of the study  is to resolve these computational difficulties.  In section~2, we review the construction of  the time independent solutions with boundary 't Hooft operators. The \textbf{KW} equations are  reduced to a Toda system. The formula of the solutions was conjectured   in a simple way by matching  boundary conditions of the half space by  Mikhaylov in \cite{vm}. In section~3, we illustrate the  construction of  the solutions precisely through an example. Then we  derive  an identity using the characteristics of Lie algebra. This identity effectively reduces   the computational difficulty of the commutation  of  operators. The another  difficulty, related to the ways from the highest  weight to a certain weight in  the fundament representation, is shown by an example. In section~4, for the  Lie algebras   of  $ADE$ type,  we find an amazing formula which  can be used to reformulate the  solutions of \textbf{KW} equations. We have checked this formula for all the solutions constructed  in  \cite{vm}. There are similar results for all minuscule representations. We also discuss this formula for none simple lattice Lie algebras.
 In the appendix, more 't~Hooft operator solutions   are  collected,  checked by different methods.

\section{Kapustin-Witten equations and the  boundary conditions}
We  take $V$ to be the half space $x^3\geq 0$ in a Euclidean space with coordinates $x^0,\cdots, x^3$.  The boundary 't~Hooft operator lies along the line $x^1=x^2=x^3=0$. In \cite{vm},  Mikhaylov reduced the Kapustin-Witten equations  to a Toda systems,  and then  conjectured a  formula of  the solutions. In this section,   we review this formula following   \cite{vm} closely to which we refer the reader for more details.
\subsection{Reduction of the \textbf{KW} equations }
For time-independent solutions, one can set $A_0=\phi_3=0$ \cite{5knots}, simplifying the \textbf{KW} equations drastically. We denote the three spacial coordinates by $x^1+ix^2=z, x^3=y$  and define the following  three operators
\begin{eqnarray}
&&{\mathcal D}_1=2\pt_{\bar z}+A_1+iA_2\,,\nnr
&&{\mathcal D}_2=\pt_y+A_3-i\phi_0\,,\nnr
&&{\mathcal D}_3=\phi_1-i\phi_2\,.
\end{eqnarray}
Then the \textbf{KW} equations (\ref{bogomol1}) take the form
\begin{eqnarray}
&&[{\mathcal D}_i,{\mathcal D}_j]=0\,,\quad i,j=1..3\,,\label{commut}\\
&&\sum_{i=1}^3[{\mathcal D}_i,{\mathcal D}_i^\dagger]=0\,.\label{moment}
\end{eqnarray}
 Eqs.(\ref{commut})  are invariant under  the complexified gauge roup $ G_{\CC}$. For this complexified gauge group,   Eq.(\ref{moment}) can be interpreted as a moment map constraint  \cite{5knots}.  Concretely, Eq.(\ref{moment}) take the form
\begin{equation}
4F_{z\bar z}+[\varphi,\varphi^\dagger]-2iD_3\phi_0=0\,,
\end{equation}
where $\varphi=\phi_1-i\phi_2$.

For the solution of Eqs.(\ref{commut}),  one can take a complex gauge transformation  in which  $A_1+iA_2=A_3-i\phi_0=0$. Then  these equations imply  that $\varphi$ is holomorphic and independent of $y$. Assuming $\varphi_0(z)$ is a  solution of Eq.(\ref{commut}), one can  apply a holomorphic gauge transformation $g(z):\mathbb{C}\rightarrow G_\CC$  to it and substitute the resulting solution into the moment map equation (\ref{moment}), then
\begin{equation}
4\pt_z\left(\pt_{\bar z}h\,h^{-1}\right)+[\varphi_0^\dagger(z),h\varphi_0(z)h^{-1}]+\pt_y\left(\pt_yh\,h^{-1}\right)=0\,,\label{fullmt}
\end{equation}
where $h=g^\dagger g$.  Let $\mathfrak{h}\subset\mathfrak{g}_\CC$ be a real Cartan subalgebra of the split real form of $\mathfrak{g}_\CC$. If  we take $g=\exp(\Psi)$ for $\Psi\in \mathfrak{h}$, this equation  reduce to
\begin{equation}
\Delta_{3d}\Psi+\fr12[\varphi_0^\dagger(z),\ex^{2\Psi}\varphi_0(z)\ex^{-2\Psi}]=0\,.\label{abelmt}
\end{equation}

In the Chevalley basis  of Lie algebra $\mathfrak{g}$, for a simple roots $\alpha_i$, denote the corresponding  raising and lowering operators by $E^\pm_i$, and the corresponding coroots by $H_i$. Then the commutation relations of these operators  are
\begin{equation}\label{cn}
 [E_i^{+}, E_{j}^{-}] = \delta_{ji}H_{j}, \quad  [H_{i}, E^{\pm}_{j}] = \pm A_{ji}E^{\pm}_{j}, \quad [H_i, H_j] =0 .
\end{equation}
The 't~Hooft operators correspond to    elements of the cocharacter lattice $\Gamma_{ch}^\vee\in{\mathfrak h}$ which is the lattice of homomorphisms ${\rm Hom}\left(\CC^*,G_\CC\right)$. Let $g(z)=\exp^{\omega\ln z}, \omega=\sum_i k_iH_i\in \Gamma_{ch}^\vee$ be such a homomorphism. Using Weyl equivalence, one can transform $\omega$ to the positive Weyl chamber such that
\begin{equation}
\mathfrak{r}_i=\alpha_i(\omega)\ge 0\,.
\end{equation}
Since the  lattice $\Gamma_{ch}^\vee$ lies inside the dual root lattice $\Gamma_r^*$,  the numbers $\mathfrak{r}_i$ are integer.

 One can take the solution of the holomorphic equations (\ref{commut}) to be of the  form
\begin{equation}
\varphi_0(z)=g(z)\varphi_1 g^{-1}(z)\,
\end{equation}
where $\varphi_1=\sum_i E^+_i$ is a representative of the principal nilpotent orbit in the algebra.
By using the commutation relations (\ref{cn}), the above formula  become
\begin{equation}
\varphi_0(z)=\sum_i z^{\mathfrak{r}_i} E^+_i\,
\end{equation}
which defines what we mean by a 't Hooft operator inserted at $z=0$ in the boundary $y=0$.  For this solution, with a real gauge transformation  $g=\exp(\Psi)$, $\Psi\in\mathfrak{h}$, the fields become
\begin{eqnarray}\label{exa}
&&A_a=-i\eps_{ab}\pt_b\Psi\,,\quad a,b=1..2\,,\nnr
&&\phi_0=-i\pt_y\Psi\,,\quad A_3=0\,,\nnr
&&\varphi=e^\Psi \varphi_0 e^{-\Psi}\,.
\end{eqnarray}
On taking a change of variables $\Psi=\fr12 \sum_{i,j}A^{-1}_{ij}H_i\psi_j$, Eq.(\ref{abelmt}) can be written in the form,
\begin{flalign}\label{td}
\sum_j A^{-1}_{sj}\Delta_{3d}\psi_j-r^{2\mathfrak{r}_s}\ex^{\psi_s}=0\,.
\end{flalign}
A convenient parameterization
\begin{equation}
\psi_i=q_i-2m_i\log r\,,\quad m_i=\mathfrak{r}_i+1\,,
\end{equation}
which brings Eq.(\ref{td}) to the scale invariant form. For the scale invariant solutions,  $q_i$ depend only on the ratio $y/r$. Setting   $y/r=\sinh\sigma$,  then  Eq.(\ref{td}) gives the Toda form\cite{He:2016prj}
\begin{equation}
\ddot{q}_i-\sum_j A_{ij}\ex^{q_j}=0\,,\label{Toda}
\end{equation}
where the dots denote derivatives with respect to $\sigma$.

 \begin{flushleft}
 \textbf{Boundary conditions}:
\end{flushleft}
To find the solutions, the boundary conditions must be fixed in order.   The boundary condition on the plane $y=0$ away from the defect is determined  by prescribing the singular behaviour of the fields \cite{5knots,Gaiotto}. In the model solution,  the gauge field  is $A_0=A_1=A_2=A_3=0$, the normal component of one form is $\phi_3=0$, and the tangent components of the one-form behave as follows
\begin{equation}
\phi_0=\fr{t_3}{y}\,, \quad \varphi=\fr{t_1-it_2}{y}\, \label{principal}
\end{equation}
where $t_i\in \mathfrak{g}_\CC$ are the images of  a principle embedding of the $\mathfrak{su}(2)$ subalgebra. This conjugacy class can be take as follows
\begin{eqnarray}
&&t_3=\fr{i}{2}\sum_i B_iH_i\,,\nnr
&&t_1-it_2=\sum_i\sqrt{B_i}E^+_i\,
\end{eqnarray}
with  $B_i=2\sum_j A^{-1}_{ij}$ \cite{Cahn}.   If $\delta^\vee\in\mathfrak{h}$ is the dual of the Weyl vector with $\alpha_i(\delta^\vee)=1$, then $t_3=i\delta^\vee$.

Let $\Delta_s$, $s=1,\dots {\rm rank}(\mathfrak{g})$, be the set of weights of the fundamental representations $\rho_s$ of the Lie algebra $\mathfrak{g}_\CC$, and $\Lambda_s$ be the highest weight. Then weight $w\in \Delta_s$ of level $n(w)$ can be represented as
\begin{equation}
w=\Lambda_s-\sum_{l=1}^{n(w)}\alpha_{j_l}\,, \quad \alpha_i\in\Delta\,.
\end{equation}
The lowest weight can be formulated as  $\tilde\Lambda_i=\Lambda_i-\sum_j n_j\alpha_j$ which relates to the height  $B_i$  as follow\,
\begin{equation}\label{bb}
 B_i= \sum_j n_j.
\end{equation}

By following \cite{5knots},  the Toda system  Eq.(\ref{Toda}) have a simple exact  solution,
\begin{equation}
q_i=-2\log\sinh\sigma+\log B_i\,.\label{model0}
\end{equation}
Then the corresponding fields in Eq.(\ref{exa}) are
\begin{equation}
A_a=i\eps_{ab}\fr{x_b}{r^2}\omega\,,\quad \phi_0=\fr{i}{2y}\sum_i B_iH_i\,,\quad \varphi=\fr{1}{y}\sum_i\left(z/\bar z\right)^{\mathfrak{r}_i/2}B_i^{1/2}E^+_i\,.
\end{equation}
This solution is singular at $r=0$.  A gauge transformation  $\tilde g=\left(\bar z/z\right)^{\omega/2}$
brings it to the form of Eq.(\ref{principal})
\begin{equation}
A_a=0\,,\quad \phi_0=\fr{i}{2y}\sum_iB_iH_i\,,\quad \varphi=\fr{1}{y}\sum_iB_i^{1/2}E^+_i\,.
\end{equation}
In order to satisfy   the boundary condition at $\sigma\rightarrow 0$,   the functions $q_i$ should approach  the model solution (\ref{model0}),
\begin{equation}
\sigma\rightarrow 0: \quad q_j=-2\log\sigma+\log B_j+\dots\,.
\end{equation}
 In the parametrization $\chi_i=\sum_j A^{-1}_{ij}q_j$ , this boundary condition can be expressed as
\begin{equation}
\sigma\rightarrow 0: \quad \ex^{-\chi_i}=\sigma^{B_i}\,\prod_k B_k^{-A^{-1}_{ik}}+\dots\,\rightarrow 0\,.\label{bc1}
\end{equation}

For $\sigma\to\infty$,  the fields must be non-singular along the line $r=0$,
\begin{equation}
\sigma\rightarrow\infty:\quad q_i=-2m_i\sigma+\log (4C_j)+{\rm O}(\ex^{-\sigma})\,,\quad m_i=\mathfrak{r}_i+1\, , \nonumber
\end{equation}
where constants $C_j$ are fixed by the boundary conditions at $\sigma=0$. The last term ${\rm O}(\ex^{-\sigma})$ is determined by  the general properties of  the open Toda systems Eq.(\ref{Toda}). In terms of variables $\chi_i$ the boundary condition is \cite{books}
\begin{equation}
\sigma\rightarrow\infty:\quad \chi_i=-2\lambda_i\sigma+\eta_i+{\rm O}(\ex^{-\sigma})\,,\label{bc2}
\end{equation}
where $\eta_i$ are functions of constants $C_j$, and $\lambda_i=\sum_j A^{-1}_{ij}m_j$.

\subsection{The Solutions}
Setting  $\chi=\sum_i\chi_iH_i$ and $\hat{\omega}=\sum_i\lambda_iH_i$,  in terms of the notations of the  previous subsection, one have
\begin{equation}
\hat{\omega}=\omega+\delta^\vee\,.
\end{equation}
 Since  $r_i=\alpha_i(\omega),\ \alpha_i(\delta^{\vee})=1$,  one have
 \begin{equation}\label{mmm}
  m_i\equiv r_i+1=\alpha_i(\hat{\omega}).
 \end{equation}

In \cite{vm}, firstly,  Milkhaylov   constructed  a solution starting from `initial values' at $\sigma\to\infty$  (\ref{bc2}).  The constants  $C_j$ can be fixed by matching the  boundary condition on the other side (\ref{bc1}).
Solution  of the open Toda system (\ref{Toda}) at time $\sigma$  is related to solution at the different time $\tau$ \cite{Kostant,Mansfield}
\begin{equation}
\ex^{-\chi_s(\sigma)}=\ex^{-\chi_s(\tau)}\langle\Lambda_s|\exp\left[(\tau-\sigma)\dot{\chi}(\tau)+\sqrt{-1}(\tau-\sigma)\sum_j\ex^{q_j(\tau)/2}(E^+_j+E^-_j)\right]|\Lambda_s\rangle\,\nonumber
\end{equation}
where $|\Lambda_s\rangle$ is the highest weight vector of unit norm in the representation $\rho_s$. By using the above formula, functions $\chi_i(\sigma)$  can be determined by  taking the limit $\tau$ to infinity and to fit the boundary conditions (\ref{bc1}),
\begin{flalign}
&\ex^{-\chi_s(\sigma)}=\lim_{\tau\to\infty}\ex^{2\lambda_s\tau-\eta_s}\langle\Lambda_s|\exp\left[2(-\tau+\sigma)\hat{\omega}+\tau\sum_j\ex^{-m_j\tau}\sqrt{-4C_j}(E^+_j+E^-_j)\right]|\Lambda_s\rangle\,\label{main2}\,.&
\end{flalign}
The following formula can be used to calculate the above limit  explicitly
\begin{equation}
\ex^{A+B}=\sum_m\int_0^1{\rm d}t_m\int_0^{t_m}{\rm d}t_{m-1}\dots\int_0^{t_2}{\rm d}t_1\ex^{(1-t_m)A}B\ex^{(t_m-t_{m-1})A}B\dots B\ex^{t_1A}\,.\label{polyakov}
\end{equation}
By choosing  operators $A=\tau\sum_j\ex^{-m_j\tau}\sqrt{-4C_j}E^-_j,  B=2(-\tau+\sigma)\hat{\omega}$,  Eq.(\ref{polyakov})  leads to
\begin{equation}
\ex^{A+B}|\Lambda\rangle=\sum_{m=0}^\infty \sum_{k=0}^m \ex^{A_k}\fr{1}{\prod_{j\ne k}(A_k-A_j)}B\dots B|\Lambda\rangle\nonumber.
\end{equation}
Upon substituting operators  $A$ and $B$, this  formula can be written in a compact form
\begin{equation}\label{e}
\ex^{-\chi_s(\sigma)}=\ex^{-\eta_s}\sum_{w\in\Delta_s}\left[\exp\left(2\sigma w(\hat{\omega})\right)\langle v_w(\hat{\omega})|v_w(\hat{\omega})\rangle (-1)^{n(w)} \prod_{l=1}^{n(w)} C_{j_l}\right],
\end{equation}
where the vector $|v_w(\hat{\omega})\rangle$ is
\begin{equation}\label{vector}
|v_w(\hat{\omega})\rangle=\sum_{\bf s} \prod_{a=1}^{n(w)} \frac{1}{w(\hat{\omega})-w_a(\hat{\omega})}E_{j_{n(w)}}^-\dots E_{j_1}^-|\Lambda\rangle\,.
\end{equation}
The notation ${\bf s}$ enumerate ways from the highest weight $\Lambda$ to a certain weight $w$,  corresponding to a sequence   $\Lambda=w_1, w_2, \dots, w_{n(w)}, w_{n(w)+1}=w\,$.

The constants $C_i$ are fixed by  matching the boundary  condition Eq.(\ref{bc1})
\begin{equation}
\sum_{w\in\Delta_s}\left[\langle v_w(\hat{\omega})|v_w(\hat{\omega})\rangle (-1)^{n(w)} \prod_{l=1}^{n(w)} C_{j_l}\right]=0\,.\nonumber
\end{equation}
In \cite{vm}, Mikhaylov made the following conjecture
\begin{equation}
C_i=\prod_{\beta_j\in\Delta_+}\left(\beta_j(\hat{\omega})\right)^{2\langle\alpha_i,\beta_j\rangle/\langle\beta_j,\beta_j\rangle}\,,\label{conj}
\end{equation}
where $\Delta_+$ is the set of positive roots. After substituting the  explicit expression of the constants $\eta_i$ in terms of $C_j$, Eq.(\ref{e}) becomes
\begin{eqnarray}\label{kwe}
 &&\ex^{-\chi_s(\sigma)}\nonumber\\
 &=&2^{-B_s}\sum_{w\in\Delta_s}\left[\exp\left(2\sigma w(\hat{\omega})\right)\,\langle v_w(\hat{\omega})|v_w(\hat{\omega})\rangle\,(-1)^{n(w)}\,\prod_{\beta_a\in\Delta_+}\left(\beta_a(\hat{\omega})\right)^{-2\langle w,\beta_a\rangle/\langle\beta_a,\beta_a\rangle}\right]\nonumber \\
 &=&\sum_{w\in\Delta_s}Q^{i}_{w}(\hat{\omega})\exp\left(2\sigma w(\hat{\omega})\right)
\end{eqnarray}
with   a Weyl invariant form $Q^{i}_{w}(\hat{\omega})$.
For the $A_n$ algebra,  this above formula   has been  proved in \cite{vm}.
Since the fundamental representations of $A_n$ are minuscule, the coefficients $Q^{i}_{w}(\hat{\omega})$ can be restored
 from the highest weight term by Weyl transformations. Then the   rewritten   formula is simple enough to check the boundary condition (\ref{bc1}) directly.

\section{Check of the boundary condition}
In the first subsection,   we refine the  factor $F_w$ in the solutions. In the second subsection, we show the  check of the boundary condition Eq.(\ref{kwe}) through an example.   In the third subsection,  we  derive  an identity which effectively simplifies  the  commutation  work  of  generators of Lie algebra.

Firstly, we summarize  the  results   in the previous  section.  The 't Hooft operator correspond to  cocharacter $\omega\in\Gamma^\vee_{ch}$.  Let $\Delta$ be the set of simple roots $\alpha_i$, and then $\alpha_i(\hat{\omega})=m_i$ with $\hat{\omega}\equiv\omega+\delta^\vee$.  $E_\alpha$ are the raising generators corresponding to the simple roots, and  then the explicit fields on the solution are
\begin{eqnarray}
&&\phi_0=-\fr{i}{2\rho}\pt_\sigma\chi(\sigma)\,,\nnr
&&\varphi=\fr{1}{r}\sum_{\alpha\in\Delta}\exp\left[\alpha(i\omega\theta+\fr12\chi(\sigma))\right]E_\alpha\,,\nnr
&&A=-i\left(\hat{\omega}+\fr12\frac{y}{\sqrt{y^2+r^2}}\pt_\sigma\chi(\sigma)\right){\rm d}\theta\, ,\nonumber
\end{eqnarray}
where $\chi(\sigma)=\sum \chi_i(\sigma)H_i$. The functions $\chi_i(\sigma)$ are conjectured in Eq.(\ref{kwe}). In order to prove this conjecture, we  need to check the following boundary condition Eq.(\ref{bc1})
\begin{equation}\label{bbc}
 \sigma\rightarrow 0: e^{-\chi_s(\sigma)}=0.
\end{equation}

For a  weight $w=\sum\limits_{i=1}^{rank(\mathfrak{g})}\lambda_i\omega_i$ in a fundament representation of  $\mathfrak{g}$, we introduce the following notations
\begin{eqnarray}\label{ewf}
  E_w &= & \textrm{exp}(2\sigma w(\hat{\omega}))(-1)^{n(w)} \nonumber\\
  W_w &=& \langle \upsilon_{w}(\hat{\omega})|\upsilon_{w}(\hat{\omega})\rangle \\
  F_w&=& \prod\limits_{\beta_{a}\in \Delta_+}(\beta_a(\hat{\omega}))^{-2\langle w,\beta_a\rangle/\langle \beta_a,\beta_a\rangle}\nonumber
\end{eqnarray}
which lead to
\begin{equation}\label{nkwe}
e^{-\chi_s(\sigma)}=2^{-B_s}\sum\limits_{w \in \Delta_s}\left[E_w\cdot W_w\cdot F_w\right].
\end{equation}

\subsection{The factor $F_w$}
We can refine the factor $F_w$ in Eq.(\ref{nkwe}) further.
For the simple root $\alpha_i$ and the fundamental weight $\omega_i$, we  have the following identities
$$\langle \omega_i,\alpha_j^{\vee}\rangle=\delta_{i,j}, \quad \alpha_i=\sum\limits_j A_{ij}\omega_j.$$
Therefore, the inner product of the positive root $\beta_a=\sum\limits_{i=1}^{rank(\mathfrak{g})}a_i\alpha_i$ is
$$\langle \beta_a,\beta_a\rangle=\sum\limits_{i=1}^{rank(\mathfrak{g})}\sum\limits_{j=1}^{rank(\mathfrak{g})}a_ia_j\langle \alpha_i,\alpha_j^{\vee}\frac{|\alpha_j|^2}{2}\rangle=\sum\limits_{i=1}^{rank(g)}\sum\limits_{j=1}^{rank(\mathfrak{g})}a_ia_jA_{ij}\frac{|\alpha_j|^2}{2}.$$
Another two factors in $F_w$ are
$$\langle w,\beta_a\rangle=\sum\limits_{i=1}^{rank(\mathfrak{g})}a_i\langle w,\alpha_i^{\vee}\frac{|\alpha_i|^2}{2}\rangle=\sum\limits_{i=1}^{rank(g)}a_i\lambda_i\frac{|\alpha_i|^2}{2}, \quad \beta_a(\hat{\omega})=\sum\limits_{i=1}^{rank(\mathfrak{g})}a_i \alpha_i(\hat{\omega})=\sum\limits_{i=1}^{rank(\mathfrak{g})}a_i m_i.$$
Substituting the above results into Eq.(\ref{ewf}), we have
\begin{equation}\label{f}
{F_w=\prod\limits_{\beta_{a}\in \Delta_+}(\beta_a(\hat{\omega}))^{-2\langle w,\beta_a\rangle/\langle \beta_a,\beta_a\rangle}
=\prod\limits_{\beta_{a}\in \Delta_+}(\sum\limits_{i=1}^{rank(\mathfrak{g})}a_i m_i)^{-2\frac{\sum\limits_{i=1}^{rank(\mathfrak{g})}a_i\lambda_i{|\alpha_i|^2}}{\sum\limits_{i=1}^{rank(\mathfrak{g})}\sum\limits_{j=1}^{rank(\mathfrak{g})}a_ia_jA_{ij}{|\alpha_j|^2}}}}.
\end{equation}
For $ADE$ groups, all the positive roots have the same length with $\langle \beta_a,\beta_a\rangle=2$. We can simply the factor $F_w$ further
\begin{equation}\label{ff}
 \boxed{F_w=\prod\limits_{\beta_{a}\in \Delta_+}(\beta_a(\hat{\omega}))^{-2\langle w,\beta_a\rangle/\langle \beta_a,\beta_a\rangle}
=\prod\limits_{\beta_{a}\in \Delta_+}(\sum\limits_{i=1}^{rank(\mathfrak{g})}a_i m_i)^{-{\sum\limits_{i=1}^{rank(\mathfrak{g})}a_i\lambda_i}}}
\end{equation}
This compact form only involves   basic dates  and simple algebraic calculation of Lie  algebra $\mathfrak{g}$, which is convenient for computer program to work on.

In section 4, we will find that there is a close relationship between the  term $ W_w$ and  term $ F_w$ for Lie algebras of  $ADE$ type, which  can be used to rewrite the solutions.
\subsection{Example: fundament representation  $\rho_1$ of $A_2$.}\label{cb}
 The highest  weight   is $\Lambda_1=[1,0]$
 \begin{equation}\label{w}
   [ 1, 0] \stackrel{\alpha_1}{\rightarrow}[-1, 1] \stackrel{\alpha_2}{\rightarrow}[ 0, -1].
 \end{equation}
There are three weights $[1,0],[-1,1],[0,-1]$ in the fundament representation $\rho_1$.  According to  Eq.(\ref{bb}), we have  $B_1=2.$
The Cartan matrix of $A_2$ is
$$A=\left(
      \begin{array}{cc}
        2 & -2 \\
        -1 & 2 \\
      \end{array}
    \right)
$$
which leads to
$$
\left(
                  \begin{array}{c}
                    \alpha_1 \\
                    \alpha_2 \\
                  \end{array}
                \right)=
                \left(
                  \begin{array}{c}
                    2\omega_1-2\omega_2 \\
                    -\omega_1+2\omega_2 \\
                  \end{array}
                \right).
$$
The positive roots  are $\Delta_+=\{\alpha_1,\alpha_2,\alpha_1+\alpha_2\}$ with lengths $|\alpha_2|^2=|\alpha_1|^2=|\alpha_1+\alpha_2|^2=2$.
 For a general weight $w=\lambda_1 \omega_1+\lambda_2 \omega_2$, using Eq.(\ref{mmm}), we have
$$
w(\hat{\omega})=(\lambda_1\omega_1+\lambda_2\omega_2)(\hat{\omega})
=(\lambda_1,\lambda_2)A_{ij}^{-1}\left(
                         \begin{array}{c}
                           \alpha_1(\hat{\omega}) \\
                           \alpha_2(\hat{\omega}) \\
                         \end{array}
                       \right)
=(\lambda_1,\lambda_2)\left(
                         \begin{array}{c}
                           \frac{2}{3} m_1+  \frac{1}{3}m_2 \\
                           \frac{1}{3}m_1+  \frac{2}{3}m_2 \\
                         \end{array}
                       \right).\\
$$

First, we calculate the factor  $F_w=\prod\limits_{\beta_{a}\in \Delta_+}(\beta_a(\hat{\omega}))^{-2\langle w,\beta_a\rangle/\langle \beta_a,\beta_a\rangle}$ in Eq.(\ref{kwe}). For the  positive roots $\alpha_1$, $\alpha_2$ in $\Delta_+$, we have
\begin{eqnarray}\label{b12}
&&\beta_1=\alpha_1:\ \ \ (\alpha_1(\hat{\omega}))^{-\langle \omega,\alpha_1^{\vee}\rangle}=m_1^{-\lambda_1}\nonumber\\
&&\beta_2=\alpha_2:\ \ \ (\alpha_2(\hat{\omega}))^{-\langle \omega,\alpha_2^{\vee}\rangle}=m_2^{-\lambda_2}
\end{eqnarray}
For the third positive  root $\beta_3=\alpha_1+\alpha_2$, we have
\begin{eqnarray*}
\langle w, \alpha_1+\alpha_2\rangle=\lambda_1+\lambda_2
\end{eqnarray*}
which leads to
\begin{equation}\label{b3}
  \beta_3=\alpha_1+\alpha_2:\ \ \ ((\alpha_1+\alpha_2)(\hat{\omega}))^{-2\langle w,\alpha_1+\alpha_2\rangle/\langle \alpha_1+\alpha_2,\alpha_1+\alpha_2\rangle}=(m_1+m_2)^{-(\lambda_1+\lambda_2)}.
\end{equation}
Combining Eq.(\ref{b12}) and Eq.(\ref{b3}), for a general weight $w=\lambda_1 \omega_1+\lambda_2 \omega_2$,  we have
\begin{equation}\label{aff}
F_{w}=\frac{1}{m_1^{\lambda_1}m_2^{\lambda_2}(m_1+m_2)^{\lambda_1+\lambda_2}}
\end{equation}
which is consistent with the formula (\ref{ff}).

Next, for each weight $w$, we calculate  terms $E_w$,  $W_w$, and $E_w \cdot W_w\cdot F_w$ in Eq.(\ref{ewf}).
For the highest weight $\Lambda$, we have
$$E^{+}_{i}|\Lambda\rangle=0,\quad H_{i}|\Lambda\rangle=\lambda_i|\Lambda\rangle.$$
The following commutation relationship  will be used  frequently
$$\langle \Lambda| E^{+}_{i} E^{-}_{i}|\Lambda\rangle= \langle \Lambda|[ E^{+}_{i},  E^{-}_{i}]+E^{-}_{i} E^{+}_{i}|\Lambda\rangle= \langle \Lambda|H_i+E^{-}_{i} E^{+}_{i}|\Lambda\rangle=\lambda_i.$$
\begin{itemize}
  \item $[1,0]$:  the level is $n([1,0])=0$. We have
$$W_{[1,0]}=\langle\Lambda|\Lambda\rangle=1,$$
and
$$E_{[1,0]}=\textrm{exp}[2\sigma([1,0])(\hat{\omega})](-1)^0={\textrm{exp}[\frac{2}{3}\sigma(2m_1+m_2)]}.$$
According to Eq.(\ref{ewf}), we get
\begin{equation}\label{f1}
F_{[1,0]}=\frac{1}{(m_1)(m_2+m_1)}.
\end{equation}
Combining the above three factors, we have
\begin{equation}\label{ewf1}
E_{[1,0]}\cdot W_{[1,0]}\cdot  F_{[1,0]}={\textrm{exp}[\frac{2}{3}\sigma(2m_1+m_2)]}\frac{1}{m_1(m_2+m_1)}.
\end{equation}

  \item $[-1,1]$: the level is $n([-1,1])=1$. We have
$$E_{[-1,1]}=\textrm{exp}[2\sigma([-1,1])(\hat{\omega})](-1)^1=-{\textrm{exp}[\frac{2}{3}\sigma(-m_1+m_2)]}.$$
According to Eq.(\ref{ewf}), we get
\begin{equation}\label{f2}
F_{[-1,1]}=\frac{m_1}{m_2}.
\end{equation}
The vector corresponding to $[-1,1]$ is
$$|\upsilon_{[-1,1]}(\hat{\omega})\rangle=\frac{1}{([1,0])(\hat{\omega})-([-1,1])(\hat{\omega})}E_{\alpha_1}^{-} |\Lambda\rangle=\frac{1}{-m_1}E_{\alpha_1}^{-} |\Lambda\rangle.$$
And the inner product of this vector is
$$W_{[-1,1]}=\langle\upsilon_{[-1,1]}(\hat{\omega})|\upsilon_{[-1,1]}(\hat{\omega})\rangle=\langle\Lambda|E_{\alpha_1}^{+}\frac{1}{-m_1}|\frac{1}{-m_1}E_{\alpha_1}^{-}|\Lambda\rangle= \frac{1}{m_1^2} \langle\Lambda|H_{\alpha_1}|\Lambda\rangle=\frac{1}{m_1^2}.$$
Combining the above three factors, we have
\begin{equation}\label{ewf2}
 E_{[-1,1]}\cdot W_{[-1,1]}\cdot F_{[-1,1]}=-{\textrm{exp}[\frac{2}{3}\sigma(-m_1+m_2)]}\frac{1}{m_1 m_2}
\end{equation}

  \item  $[0,-1]$: the level is $n([0,-1])=2$. We have
$$E_{[0,-1]}={\textrm{exp}[-\frac{2}{3}\sigma(m_1+2m_2)]}(-1)^2.$$
According to Eq.(\ref{ewf}), we get
\begin{equation}\label{f3}
F_{[0,-1]}={m_2}{(m_1+m_2)}.
\end{equation}
The vector corresponding to $[-1,1]$ is
\begin{eqnarray*}
|\upsilon_{[0,-1]}(\hat{\omega})\rangle &=&\frac{1}{w(\hat{\omega})-w_2(\hat{\omega})}\cdot\frac{1}{w(\hat{\omega})-w_1(\hat{\omega})}E_{\alpha_2}^{-} E_{\alpha_1}^{-} |\Lambda\rangle\\
&=&\frac{1}{-m_2}\cdot\frac{1}{-m_1-m_2}E_{\alpha_2}^{-} E_{\alpha_1}^{-} |\Lambda\rangle.
\end{eqnarray*}
The conjugate vector is
$$\langle \upsilon_{[0,-1]}(\hat{\omega})|=\langle\Lambda|E_{\alpha_1}^{+} E_{\alpha_2}^{+}\frac{1}{m_2(m_1+m_2)}.$$
And the inner product  is
\begin{eqnarray*}
 W_{[0,-1]} &=& \langle\upsilon_{[0,-1]}(\hat{\omega})|\upsilon_{[0,-1]}(\hat{\omega})\rangle \\
   &=&\frac{1}{(m_2(m_1+m_2))^2} \langle\Lambda|E_{\alpha_1}^{+} E_{\alpha_2}^{+}E_{\alpha_2}^{-} E_{\alpha_1}^{-} |\Lambda\rangle\\
     &=&  \frac{1}{(m_2(m_1+m_2))^2}.
   \end{eqnarray*}
Combining the  above results, we have
\begin{equation}\label{ewf3}
 E_{[0,-1]}\cdot W_{[0,-1]}\cdot F_{[0,-1]}={\textrm{exp}[\frac{2}{3}\sigma(m_1+2m_2)]}\frac{1}{m_2(m_1+ m_2)}.
\end{equation}
\end{itemize}

Substituting   Eqs.(\ref{ewf1}), (\ref{ewf2}), and (\ref{ewf3}) to the formula (\ref{ewf}), we have
\begin{eqnarray*}
  e^{-\chi_1(\sigma)} &=& 2^{-2}(E_{[1,0]}\cdot W_{[1,0]}\cdot  F_{[1,0]} +E_{[-1,1]}\cdot W_{[-1,1]}\cdot F_{[-1,1]}+E_{[0,-1]}\cdot W_{[0,-1]}\cdot F_{[0,-1]})\\
   &=& \frac{1}{4}(\frac{{\textrm{exp}[\frac{2}{3}\sigma(2m_1+m_2)]}}{m_1(m_2+m_1)}-\frac{{\textrm{exp}[\frac{2}{3}\sigma(-m_1+m_2)]}}{m_1 m_2}+\frac{{\textrm{exp}[\frac{2}{3}\sigma(m_1+2m_2)]}}{m_2(m_1+ m_2)})
\end{eqnarray*}
which is consistent with the result  in \cite{vm}. It is easy to check that the above formula satisfy the boundary condition Eq.(\ref{bbc})
$$\sigma\rightarrow 0: \qquad e^{-\chi_1(\sigma)}=0.$$

 From the above derivations, we find that  the calculation of the commutation  of  operators in $W_w$ is a boring job for checking the boundary condition Eq.(\ref{kwe}). In a similar situation, it is  unrealistic   for a personal computer to work out  the inner product of a state, with more than ten Virasoro operators $L_n$ acting on the highest weight state, in finite time. There is another computation difficulty in $W_w$. In this example there is only one way reaching a weight from the highest weight. With the  rank of Lie algebra $\mathfrak{g}$ increasing , the number of weights as well as  the number of  ways reaching a weight increase rapidly. As a result, the calculation work  increases rapidly if realizing  the commutation of  operators   directly. This will become clear in an example in the next subsection.

\subsection{The vanishing factor}
 In this subsection we  derive an identity  to  reduce  the commutation  work  of generators of Lie algebra  in the factor $W_w$.
For the highest weight $\Lambda=\Sigma_a \lambda_a \omega_a$, according to the commutation  relations (\ref{cn}), we have the following basic identity,
 \begin{eqnarray}\label{le0}
  E_{a}^{+}E_{j_n}^{-} E_{j_{n-1}}^{-}\cdots E_{j_1}^{-} |\Lambda\rangle &=& (\delta_{a,j_n}H_{a}+E_{j_n}^{-}E_{a}^{+} )E_{j_{n-1}}^{-}\cdots E_{j_1}^{-} |\Lambda\rangle \nonumber\\
    &=&\sum_{i=1}^{n}E_{j_n}^{-} E_{j_{n-1}}^{-}\cdots \delta_{a,j_i}H_{a} \hat{E}_{j_{i}}^{-}E_{j_{i-1}}^{-}\cdots E_{j_1}^{-}  |\Lambda\rangle\\
    &=& \sum_{i=1}^{n}\delta_{a,j_i}(\lambda_a-(\sum_{i=1}^{i-1}A_{j_l,a}))E_{j_n}^{-} E_{j_{n-1}}^{-}\cdots \hat{E}_{j_{i}}^{-}E_{j_{i-1}}^{-}\cdots E_{j_1}^{-} |\Lambda\rangle\nonumber
 \end{eqnarray}
where the hat means omitting the corresponding term. In a  special case,
\begin{eqnarray}\label{le}
  E_{i}^{+}(E_{i}^{-})^n |\Lambda\rangle
    &=&(H_i+ E_{i}^{-}E_{i}^{+})(E_{i}^{-})^{n-1} |\Lambda\rangle\nonumber\\
  &=& \sum_{i=1}^{n-1} (E_{i}^{-})^{l} H_i (E_{i}^{-})^{n-1-l} |\Lambda\rangle\nonumber \\
    &=&\sum_{i=1}^{n-1}(E_{i}^{-})^{l}(\lambda_i-(n-1-l)A_{ii})(E_{i}^{-})^{n-1-l}|\Lambda\rangle \\
       &=&n(\lambda_i-(n-1))(E_{i}^{-})^{n-1}|\Lambda\rangle.\nonumber
\end{eqnarray}
We can generalize the identity  (\ref{le0})  further. The following identity is one of the main results we get in this paper.
\begin{Theorem}
For the highest weight $\Lambda=\Sigma_a \lambda_a \omega_a$, we have
\begin{equation*}\label{pro}
  E_{i}^{+}(E_{i}^{-})^n \prod_{i=1}^{m}E_{j_b}^{-}) |\Lambda\rangle
  =n(\lambda_i-(n-1)-\sum_{b=1}^{m}A_{j_b,i}) (E_{i}^{-})^{n-1} \prod_{i=1}^{m}E_{j_b}^{-})|\Lambda\rangle+ (E_{i}^{-})^{n} E_{i}^{+}\prod_{i=1}^{m}E_{j_b}^{-})|\Lambda\rangle
\end{equation*}
\end{Theorem}
\begin{flushleft}
\textrm{\textbf{Proof}}: According to   Eq.(\ref{le}), we have
\end{flushleft}
\begin{eqnarray*}
  L.H.S&=& \sum_{a=0}^{n-1} (E_{i}^{-})^{a} H_i (E_{i}^{-})^{n-1-a} \prod_{i=1}^{m}E_{j_b}^{-})|\Lambda\rangle+ (E_{i}^{-})^{n} E_{i}^{+}\prod_{i=1}^{m}E_{j_b}^{-})|\Lambda\rangle\\
    &=&\sum_{a=0}^{n-1} (E_{i}^{-})^{a} (\lambda_i-(n-1-a)A_{ii}-\sum_{b=1}^{m}A_{j_b,i}) (E_{i}^{-})^{n-1-a} \prod_{i=1}^{m}E_{j_b}^{-})|\Lambda\rangle\\
    &&\quad \quad \quad\quad + (E_{i}^{-})^{n} E_{i}^{+}\prod_{i=1}^{m}E_{j_b}^{-})|\Lambda\rangle \\
       &=&n(\lambda_i-(n-1)-\sum_{b=1}^{m}A_{j_b,i}) (E_{i}^{-})^{n-1} \prod_{i=1}^{m}E_{j_b}^{-})|\Lambda\rangle+ (E_{i}^{-})^{n} E_{i}^{+}\prod_{i=1}^{m}E_{j_b}^{-})|\Lambda\rangle.
\end{eqnarray*}
\begin{flushright}
 \textbf{ Q.E.D}
\end{flushright}
When $n=0$, this formula reduce to Eq.(\ref{le0}). When $m=0$, we recover Eq.(\ref{le}).
An important fact that we  find  is  that the following  factor
\begin{equation}\label{im}
  \boxed{(\lambda_i-(n-1)-\sum_{b=1}^{m}A_{j_b,i})}
\end{equation}
vanish  from time to time. When this factor is zero,  the first  term on the right hand side of the formula in  {Proposition \ref{pro}} can be omitted,    decreasing  the commutation   work  of   operators in $W_w$  greatly.

\begin{figure}
  \begin{center}
    \includegraphics[width=6in]{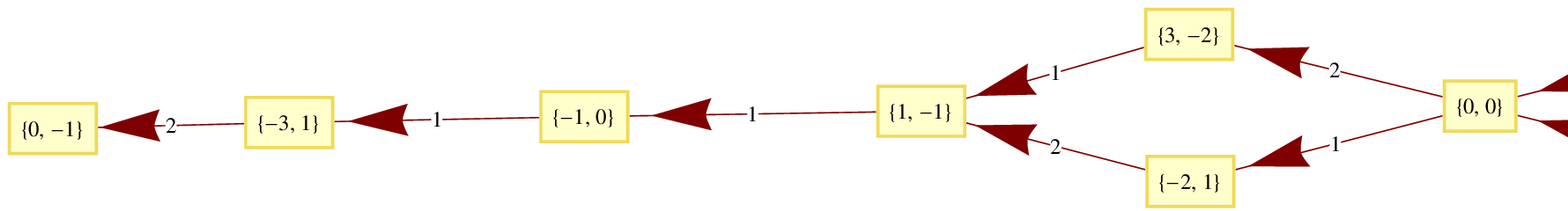}
  \end{center}
  \caption{Weights  in the fundament representation $\rho_2$ of $G_2$. The number $i$ on the arrow stand for $-\alpha_i$. }
  \label{G2-01}
\end{figure}

Before illustrating the vanishing property of the  factor (\ref{im}), we introduce a fact which is helpful in the  practical computation.
\begin{Theorem}{\label{pro2}}
 $\Lambda_i=[0,\cdots,1,\cdots,0]$ is the highest weight of the  fundament representation $\rho_i$. We introduce   the following state
$$|\nu_w(\hat{\omega})\rangle=f(E^{-}_{*})E^{-}_{j}|\Lambda_i\rangle,\, i\neq j$$
where $f$ is a polynomial function of the generators of Lie algebra $\mathfrak{g}$. For arbitrary states $\langle g(E^{+}_{*}) |$, we have
$$\langle g(E^{+}_{\ast}) |\nu_w(\hat{\omega})\rangle\equiv0.$$
\end{Theorem}
\begin{flushleft}
\textrm{\textbf{Proof}}: First, we commutate  all these operators in $f(E^{-}_{*})$ sequentially to the left side of all $E^{+}_{\ast}$ in $g(E^{+}_{\ast})$.  According to  the following identity
$$H_{k}E_{k_{n}}^{-}\cdots E_{k_1}^{-}  |\Lambda\rangle=c_{k}E_{k_{n}}^{-}\cdots E_{k_1}^{-}  |\Lambda\rangle,$$
operators $H_{k}$,  appearing in the  commutation   $[E^{+}_{k},E^{-}_{k}]$,  can be seen as a undetermined  constants $c_k$. Finally,   operator $E^{-}_{*}$ annihilate the lowest weight sate $\langle\Lambda|$. Then  only the   operators $E^{+}_{\ast}$  and $E^{-}_{j}$ are left. If  no operator   $E^{+}_{j}$ is left acting  on $E^{-}_{j}|\Lambda_i\rangle$,  the operator  $E^{-}_{j}$ will  commutate  all the operators $E^{+}_{*}$   and annihilate the state $\langle\Lambda|$ which leads to the conclusion.  If   at least one   $E^{+}_{j}$ is left,  we have
$$\langle g(E^{+}_{\ast}) |\nu_w(\hat{\omega})\rangle=\langle\cdots E^{+}_{j}E^{-}_{j}|\Lambda_i\rangle=\langle\cdots (H_j+E^{-}_{j}E^{+}_{j})|\Lambda_i\rangle=0.$$
where $H_j|\Lambda_i\rangle=0$ because of $i\neq j$ and  $ E^{+}_{j} $ annihilate the highest weight state $|\Lambda_i\rangle$.
\end{flushleft}
\begin{flushright}
 \textbf{ Q.E.D}
\end{flushright}
\begin{flushleft}

Next, we give an example to illustrate the vanishing property of   factor (\ref{im}).

\textrm{\textbf{Example}}: As shown in Fig.(\ref{G2-01}), there are  four paths reaching weight $[-3,1]$ from the highest  weight $\Lambda=[0,1]$. The path that will be handed by us  is
\end{flushleft}
\begin{eqnarray*}
&&[ 0, 1] \stackrel{\alpha_2}{\rightarrow}[3, -1] \stackrel{\alpha_1}{\rightarrow}[ 1, 0]\stackrel{\alpha_1}{\rightarrow}[-1, 1] \stackrel{\alpha_1}{\rightarrow}[ -3, 2]\stackrel{\alpha_2}{\rightarrow}[0, 0] \\
&&\qquad\qquad \qquad\qquad\qquad \qquad\stackrel{\alpha_1}{\rightarrow}[ -2, 1] \stackrel{\alpha_2}{\rightarrow}[1, -1] \stackrel{\alpha_1}{\rightarrow}[ -1, 0]\stackrel{\alpha_1}{\rightarrow}[-3, 1].
\end{eqnarray*}

The Cartan matrix of $G_2$ is
$$A=\left(
      \begin{array}{cc}
        2 & -1 \\
        -3 & 2 \\
      \end{array}
    \right).
$$
We  calculate the  following inner product which is the denominator of  $W_{[-3,1]}$. Using proposition 1, performing  the action of  operators $E^{+}_{\ast}$  sequentially,  we have
\begin{eqnarray*}
 W^{'}_{[-3,1]}  &=& \langle\Lambda|E^{+}_{2}(E^{+}_{1})^3E^{+}_{2}E^{+}_{1}E^{+}_{2}(E^{+}_{1})^2| (E^{-}_{1})^2E^{-}_{2}E^{-}_{1}E^{-}_{2}(E^{-}_{1})^3E^{-}_{2}|\Lambda\rangle\\
   &=&\langle\Lambda|E^{+}_{2}(E^{+}_{1})^3E^{+}_{2}E^{+}_{1}E^{+}_{2}E^{+}_{1}\{\underbrace{(-2-2(2A_{21}+4A_{11}))}_0E^{-}_{1}E^{-}_{2}E^{-}_{1}E^{-}_{2}(E^{-}_{1})^3E^{-}_{2}\\ &&+(E^{-}_{1})^2E^{-}_{2}\underbrace{(-2A_{21}-3A_{11})}_0 E^{-}_{2}(E^{-}_{1})^3E^{-}_{2}\\
   &&+(E^{-}_{1})^2E^{-}_{2}E^{-}_{1}E^{-}_{2}\underbrace{(-3\cdot 2-3A_{21})}_3(E^{-}_{1})^2E^{-}_{2}\}|\Lambda\rangle
\end{eqnarray*}
In this formula,  the first two terms within the braces are omitted because of the zero factor.  The third term is
\begin{eqnarray*}
 &&W^{'}_{[-3,1]}\\
  &=& 3\langle\Lambda|E^{+}_{2}(E^{+}_{1})^3E^{+}_{2}E^{+}_{1}E^{+}_{2}\{ \underbrace{(-2-2(3A_{21}+3A_{11}))}_4 E^{-}_{1}E^{-}_{2}E^{-}_{1}E^{-}_{2}(E^{-}_{1})^2E^{-}_{2}\\
  &&+(E^{-}_{1})^2 E^{-}_{2}\underbrace{(-2A_{21}-2A_{11})}_2E^{-}_{2}(E^{-}_{1})^2E^{-}_{2}+(E^{-}_{1})^2E^{-}_{2}E^{-}_{1}E^{-}_{2}\underbrace{(-2-2A_{21})}_4E^{-}_{1}E^{-}_{2}\}|\Lambda\rangle
  \\ &=&3(4W^{'}_{[-3,1]_1}+2W^{'}_{[-3,1]_2}+4W^{'}_{[-3,1]_3})
\end{eqnarray*}
where we denote the three none zero  terms as $W^{'}_{[-3,1]_1}$,$W^{'}_{[-3,1]_2}$, $W^{'}_{[-3,1]_3}$, respectively. For the first one, we have
\begin{eqnarray*}
  W^{'}_{[-3,1]_1} &=& \langle\Lambda|E^{+}_{2}(E^{+}_{1})^3E^{+}_{2}E^{+}_{1}E^{+}_{2}| E^{-}_{1}E^{-}_{2}E^{-}_{1}E^{-}_{2}(E^{-}_{1})^2E^{-}_{2}|\Lambda\rangle\\
   &=& \langle\Lambda|E^{+}_{2}(E^{+}_{1})^3E^{+}_{2}E^{+}_{1}\{ E^{-}_{1}\underbrace{(-3A_{12}-2A_{22}+\lambda_2)}_0E^{-}_{1}E^{-}_{2}(E^{-}_{1})^2E^{-}_{2}\\
  && +E^{-}_{1}E^{-}_{2}E^{-}_{1}\underbrace{(-2A_{12}-A_{22}+\lambda_2)}_1(E^{-}_{1})^2E^{-}_{2}\}|\Lambda\rangle\\
   &=&\langle\Lambda|E^{+}_{2}(E^{+}_{1})^3E^{+}_{2}\{ \underbrace{(-2A_{21}-3A_{11})}_0E^{-}_{2}(E^{-}_{1})^3E^{-}_{2}\\
   &&\qquad\qquad\qquad\qquad+E^{-}_{1}E^{-}_{2}\underbrace{(-3\cdot2-3A_{21})}_3(E^{-}_{1})^2E^{-}_{2}|\Lambda\rangle\\
  &=&3\langle\Lambda|E^{+}_{2}(E^{+}_{1})^3\{ E^{-}_{1}\underbrace{(-2A_{12}-A_{22}+\lambda_2)}_1(E^{-}_{1})^2E^{-}_{2}+\underbrace{E^{-}_{1}E^{-}_{2}(E^{-}_{1})^2E^{-}_{2}\}}_{0(Proposition\,\ref{pro2})}|\Lambda\rangle\\
  &=&3\langle\Lambda|E^{+}_{2}(E^{+}_{1})^3| (E^{-}_{1})^3E^{-}_{2}|\Lambda\rangle \\
  &=&3\cdot 36
\end{eqnarray*}
As expected, the factor $(\lambda_i-(n-1)-\sum_{b=1}^{m}A_{j_b,i})$ becomes zero frequently. This vanishing property  reduces  much computation work. For the second term, we have,
\begin{eqnarray*}
  W^{'}_{[-3,1]_1} &=& \langle\Lambda|E^{+}_{2}(E^{+}_{1})^3E^{+}_{2}E^{+}_{1}E^{+}_{2}| (E^{-}_{1})^2E^{-}_{2}E^{-}_{2}(E^{-}_{1})^2E^{-}_{2}|\Lambda\rangle\\
   &=& \langle\Lambda|E^{+}_{2}(E^{+}_{1})^3E^{+}_{2}E^{+}_{1}\{ (E^{-}_{1})^2\underbrace{(-2-2(2A_{12}+A_{22})+2\lambda_2)}_0E^{-}_{2}(E^{-}_{1})^2|\Lambda\rangle\\
  &=&0
\end{eqnarray*}
For the third one, we have
\begin{eqnarray*}
  W^{'}_{[-3,1]_3} &=& \langle\Lambda|E^{+}_{2}(E^{+}_{1})^3E^{+}_{2}E^{+}_{1}\{ (E^{-}_{1})^2\underbrace{(-2A_{12}-2A_{22}+\lambda_2)}_{-1}E^{-}_{1}E^{-}_{2}E^{-}_{1}E^{-}_{2}\\
  &&+(E^{-}_{1})^2E^{-}_{2}E^{-}_{1}\underbrace{(-A_{12}-A_{22}+\lambda_2)}_0E^{-}_{1}E^{-}_{2}|\Lambda\rangle\\
   &=& -\langle\Lambda|E^{+}_{2}(E^{+}_{1})^3E^{+}_{2}\{ \underbrace{(-3\cdot2-3(2A_{21}+A_{11}))}_6(E^{-}_{1})^2E^{-}_{2}E^{-}_{1}E^{-}_{2}\\
  &&+(E^{-}_{1})^3E^{-}_{2}\underbrace{(-A_{12})}_3E^{-}_{2}|\Lambda\rangle\\
  &=& -\langle\Lambda|E^{+}_{2}(E^{+}_{1})^3\{ 6(E^{-}_{1})^2\underbrace{(-A_{12}-A_{22}+\lambda_2)}_0E^{-}_{1}E^{-}_{2}+6\underbrace{(E^{-}_{1})^2E^{-}_{2}E^{-}_{1}\lambda_{2}}_{ 0(Proposition \, \ref{pro2})}\\
  &&+3(E^{-}_{1})^3\underbrace{(-2+2\lambda_2)}_0E^{-}_{2}|\Lambda\rangle\\
  &=&0
\end{eqnarray*}

Combining all the above results, the inner product  is
\begin{eqnarray}\label{g2w}
   W^{'}_{[-3,1]} &=& 3(4W^{'}_{[-3,1]_1}+2W^{'}_{[-3,1]_2}+4W^{'}_{[-3,1]_3})\nonumber \\
   &=& 3(4\cdot 3\cdot 36+2\cdot 0+ 4\cdot 0) \\
   &=& 36\cdot 36\nonumber
\end{eqnarray}

In the process of above computation,  the term $(\lambda_i-(n-1)-\sum_{b=1}^{m}A_{j_b,i})$ is to be  zero frequently. After  performing  many  examples, we find it is a common phenomenon. By virtue of this vanishing factor, the computational efficiency is improved remarkably and the computation of the factor $W_w$ in $e^{-\chi_s(\sigma)}$ is simplified.

\begin{figure}
  \begin{center}
    \includegraphics[width=6in]{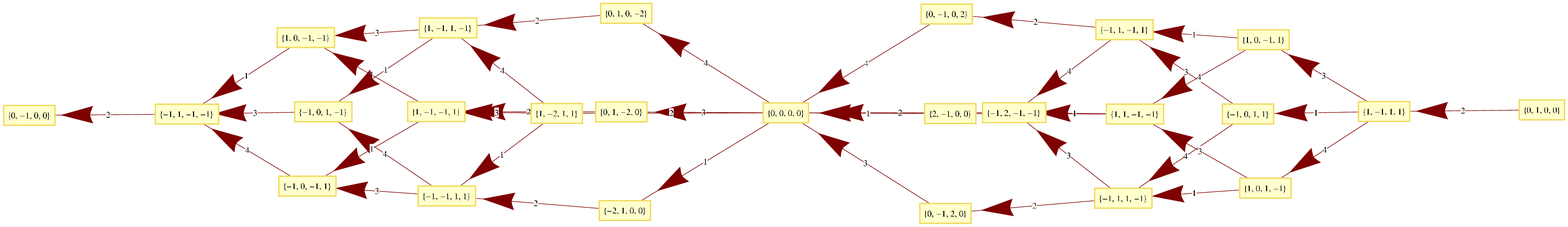}
  \end{center}
  \caption{Weights in the fundament representation $\rho_2$ of  $D_4$. Each branch node increase the number of paths  $\mathrm{\mathbf{s}}$. There are ten paths reaching  weight $[0,0,0,0]$ from the highest weight to define the vector  $|\upsilon_{[0,0,0,0]}(\hat{\omega})\rangle$.  Twenty-five  vectors $|\upsilon_{w}(\hat{\omega})\rangle$  need to be considered to compute $e^{-\chi_2}$.}
  \label{D4-0100}
\end{figure}

Unfortunately, there is another computation difficulty  pointed out at the end of Section \ref{cb}. To define the vector $|\upsilon_{w}(\hat{\omega})\rangle$, it is necessary  to consider all the ways $\mathrm{\mathbf{s}}$ reaching  $w$ from the highest weight state. As shown in Fig.(\ref{D4-0100}), each branch node increase the number of paths. There are ten paths reaching weight $[0,0,0,0]$ from the highest weight state to define the  vector  $|\upsilon_{[0,0,0,0]}(\hat{\omega})\rangle$.  With the rank of $\mathfrak{g}$ rising, the number of paths reaching a weight from the highest  weight    increase rapidly,  as well as the number of weights. Twenty-five  vectors $|\upsilon_{w}(\hat{\omega})\rangle$  need to be considered  to compute $e^{-\chi_2}$. Note that we record the inner product of the vector $|\upsilon_{[-3,1]}(\hat{\omega})\rangle$ in one page. But for the vector $|\upsilon_{[0,-1]}(\hat{\omega})\rangle$, we need more than twenty pages to record the whole calculation  process.  For these weights in the fundamental representation of   $G_2$ theory,  we can compute the factors $W_w$ by hand, but   it is unrealistic to compute the factors $W_w$ by hand for Lie algebra of higher  rank.  In fact, it is even difficult for personal  computer to work out the factor $e^{-\chi_2}$ of the  $D_4$ theory. However, in the next section, we will find another construction of the solutions of \textbf{KW} equations for semisimple Lie algebras of $ADE$ type.  This new formula of solutions   does not involve the factor  $W_w$. Thus  it   bypass the computational difficulties contained in the factor $W_w$.

\section{Construction of solutions}\label{solution}
In this section, we propose an amazing formula  that  can be used to reformulate the solutions of \textbf{KW} equations for the Lie algebras of $ADE$ type and the minuscule representation. This new formula   not only avoid  computing the commutation of operators   but also avoid the difficulty related to the number of  paths $\textrm{\textbf{s}}$ in the definition of the vector $|\upsilon_{w}(\hat{\omega})\rangle$. We  give an example in this section and more results in the  appendix to support our proposal. Unfortunately, there are no simple   rules of the solutions  for none simple lattice Lie algebras.
\subsection{$ADE$ groups and the minuscule representations}\label{solution1}
According to Eq.(\ref{vector}), for a weight $w=\Lambda_s-\sum\limits_{l=1}^{n(w)}\alpha_{j_l},\ \  \alpha_{j_l} \in \Delta$ in the fundament representation  $\rho_s$,  the  vector  $|\upsilon_{w}(\hat{\omega})\rangle$  is
$$|\upsilon_{w}(\hat{\omega})\rangle=\sum\limits_\textrm{\textbf{s}}\prod\limits_{a=1}^{n(w)}\frac{1}{w(\hat{\omega})-w_a(\hat{\omega})}
E_{j_{n(w)}}^-\cdots E_{j_1}^-|\Lambda_s\rangle.$$
Let us consider the  term
\begin{equation}\label{tt}
  \langle \upsilon_{\omega}(\hat{\omega})|\upsilon_{\omega}(\hat{\omega})\rangle
\prod\limits_{\beta_{a}\in \Delta_+}(\beta_a(\hat{\omega}))^{-2\langle w,\beta_a\rangle/\langle \beta_a,\beta_a\rangle}.
\end{equation}
 We have the following  conjecture which can simplify  the construction of the  solutions of \textbf{KW} equations.
\begin{Definition}\label{c2} For a  weight $w\in \Delta_s$ in the fundament representation  $\rho_s$ of the simple-laced Lie algebras $ (A_n, D_n, E_6,E_7,E_8)$, according to  Eq.(\ref{ff}),  we have
\begin{equation}\label{ab}
  {F_w=\prod\limits_{\beta_{a}\in \Delta_+}(\beta_a(\hat{\omega}))^{-2\langle w,\beta_a\rangle/\langle \beta_a,\beta_a\rangle}
=\prod\limits_{\beta_{a}\in \Delta_+}(\sum\limits_{i=1}^{rank(\mathfrak{g})}a_i m_i)^{-{\sum\limits_{i=1}^{rank(\mathfrak{g})}a_i\lambda_i}}}=\frac{A_w} {B_w},
\end{equation}
where the numerator $A_w$ and  denominator $B_w$  have no common factor, with variables $m_i$. The sequences $\lambda-k_i\alpha_{n_i} , k_i\in [0,\cdots,n]$   along the simple root $\alpha_{n_i}$ are elements  in the weight space $\Delta_s$,  while $\lambda+\alpha_{n_i}, \lambda-(n+1)\alpha_{n_i}$ do not belong to  the weight space. If $w\neq \lambda-k_i\alpha_{n_i}, k_i\in [1,\cdots,n-1]$,  it is  conjectured that
\begin{equation}\label{cc}
  W_w\cdot F_w=\langle \upsilon_{w}(\hat{\omega})|\upsilon_{w}(\hat{\omega})\rangle
\prod\limits_{\beta_{a}\in \Delta_+}(\beta_a(\hat{\omega}))^{-2\langle w,\beta_a\rangle/\langle \beta_a,\beta_a\rangle} =\frac{1}{A_w \cdot B_w}
\end{equation}
which means
\begin{equation}\label{cccc}
  W_w=\langle \upsilon_{w}(\hat{\omega})|\upsilon_{w}(\hat{\omega})\rangle
 =\frac{1}{(A_w)^2}.
\end{equation}
\end{Definition}
The  terms $A_w$ and   $B_w$  can be calculated  by simple algebraic relations  and do not involve the computational difficulties in $W_w=\langle \upsilon_{w}(\hat{\omega})|\upsilon_{w}(\hat{\omega})\rangle$.
According to the conjecture, all the weight which  are only in  a string of two elements along a simple root satisfy Eq.(\ref{ab}). And the first weight and  last weight in a none two elements string also satisfy Eq.(\ref{ab}).

We reanalyze the  example in Section \ref{cb} to illustrate the conjecture  \ref{c2}.
\begin{flushleft}
\textbf{Example}: fundamental representation $\rho_1$ with the highest weight $[1,0]$ of $A_2$.
\end{flushleft}
\begin{itemize}
  \item  $[1,0]$:  according to Eq.(\ref{f1}), we have
$$F_{[1,0]}=\frac{1}{(m_1)(m_2+m_1)}$$
which implies $A_w=1$ and $B_w=(m_1)(m_2+m_1)$. Using  Eq.(\ref{cc}), we get
\begin{equation}
 W_{[1,0]}\cdot  F_{[1,0]}=\frac{1}{A_w\cdot B_w}=\frac{1}{m_1(m_2+m_1)}.\nonumber
\end{equation}
  \item $[-1,1]$: according to Eq.(\ref{f2}), we have
$$F_{[-1,1]}=\frac{m_1}{m_2}.$$
which implies $A_w=m_1$ and $B_w=m_2$. Using Eq.(\ref{cc}), we get
\begin{equation}
  W_{[-1,1]}\cdot F_{[-1,1]}=\frac{1}{A_w\cdot B_w}=\frac{1}{m_1 m_2}.\nonumber
\end{equation}
  \item $[0,-1]$: according to Eq.(\ref{f3}), we have
$$F_{[0,-1]}={m_2}{(m_1+m_2)}.$$
which implies $A_w={m_2}{(m_1+m_2)}$ and $B_w=1$. Using Eq.(\ref{cc}), we get
\begin{equation}
 W_{[0,-1]}\cdot F_{[0,-1]}=\frac{1}{A_w\cdot B_w}=\frac{1}{m_2(m_1+ m_2)}.\nonumber
\end{equation}
\end{itemize}
The terms $W_{[1,0]}\cdot  F_{[1,0]}$, $W_{[-1,1]}\cdot F_{[-1,1]}$ and $W_{[0,-1]}\cdot F_{[0,-1]}$ are all consistent with the results discussed in Section \ref{cb}.

For some fundamental representations, such as $\rho_2$ of  $D_4$ as shown in Fig.(\ref{D4-0100}), the weight $[0,\cdots,0]$ is   the  only weight not in  a string of two elements along a simple root.
For these cases,  $F_w=1$ in Eq.(\ref{ewf}). One would speculate that $A_w=B_w$ which means $W_w\cdot F_w=\frac{1}{A_w^2}$. However, this naive guess is not collect. A counterexample, $W_{[0,1,0,0]}\cdot F_{[0,1,0,0]}$ in  the  fundamental representation $\rho_2$ of $D_4$,  is given in Appendix \ref{B}.

When the weight $[0,\cdots,0]$ is   the  only weight not in  a string of two elements along a simple root in the weight space $ \Delta_s$,  we can reformulate $e^{-\chi_s(\sigma)}$ using the boundary condition Eq.(\ref{bc1}).   According to this boundary condition, we have
$$\sum_w W_{w}\cdot  F_{w}|_{{\sigma=0}}=0.$$
This formula implies
\begin{equation}\label{cs}
W_{[0,\cdots,0]}\cdot  F_{[0,\cdots,0]}=-\sum_{w^{'}} W_{w^{'}}\cdot  F_{w^{'}}=-\sum_{w^{'}} \frac{1}{A_{w^{'}} \cdot B_{w^{'}}}
\end{equation}
where $w^{'}$ denotes the exclusion of $[0,\cdots,0]$.
Thus, we can construct $e^{-\chi_s(\sigma)}$ as follows
\begin{Theorem}\label{p3}
For the simple-laced Lie algebras $ (A_n, D_n, E_6,E_7,E_8)$, if  $[0,0,\cdots,0]$ is  the only weight not in  a string of two elements along a simple root in  the fundament representation $\rho_s$,
using Eq.(\ref{cs}), we have
\begin{equation}\label{nkwe2}
e^{-\chi_s(\sigma)}=2^{-B_s}(\sum\limits_{w\in \Delta_s^{'}}\left[E_w\cdot \frac{1}{A_w \cdot B_w}\right]+W_{[0,\cdots,0]}\cdot  F_{[0,\cdots,0]})=2^{-B_s}\sum\limits_{w\in \Delta_s^{'}}\frac{1}{A_w \cdot B_w}\left[ E_w- 1\right]\nonumber
\end{equation}
where ${A_w}$ and  ${B_w}$ are defined in Eq.(\ref{ab}) and  $\Delta_s^{'}$  denotes the exclusion of the weight $[0,\cdots,0]$ in $\Delta_s$.
\end{Theorem}
Examples of solutions using the above formula are given in  Appendix \ref{B}.

For the minuscule representations, all the strings are two terms long in the weight spaces, with   the fundamental weight as  the highest weight.   The following table is a complete list of minuscule fundamental weights for simple Lie algebras  \cite{Green}.

\begin{center}
\begin{tabular}{cc}
\multicolumn{2}{c}{ Minuscule fundamental weights for simple Lie algebras\footnote{cite??} }\\ \hline
Type   &   \{$i$: $\omega_i$ is minuscule \} \\  \hline
$A_l$  & $ 1,2,\cdots, l$\\
$B_l$  & l \\
$C_l$  & 1  \\
$D_l$  & $1,l-1,l$  \\
 $E_6$   & 1,5 \\
 $E_7$   & 6 \\
 $E_8$   &  none\\
  $F_4$  &  none \\
 $G_2$   &  none\\  \hline
\end{tabular}
\end{center}
For the minuscule  representations, we have the following conjecture
\begin{Definition}\label{mmd} For a  weight $w\in \Delta_s$ in the minuscule representation  $\rho_s$, according to  Eq.(\ref{f}),  we have
\begin{equation}\label{abm}
  F_w=\prod\limits_{\beta_{a}\in \Delta_+}(\beta_a(\hat{\omega}))^{-2\frac{\langle w,\beta_a\rangle}{\langle \beta_a,\beta_a\rangle}}
=\prod\limits_{\beta_{a}\in \Delta_+}(\sum\limits_{i=1}^{rank(\mathfrak{g})}a_i m_i)^{-2\frac{\sum\limits_{i=1}^{rank(\mathfrak{g})}a_i\lambda_i{|\alpha_i|^2}}{\sum\limits_{i=1}^{rank(\mathfrak{g})}\sum\limits_{j=1}^{rank(\mathfrak{g})}a_ia_jA_{ij}{|\alpha_j|^2}}}=\frac{A_w} {B_w}
\end{equation}
where the numerator $A_w$ and denominator $B_w$  have no common factor, with variables $m_i$. We have the following   conjecture
\begin{equation}\label{ccm}
  W_w\cdot F_w=\langle \upsilon_{w}(\hat{\omega})|\upsilon_{w}(\hat{\omega})\rangle
\prod\limits_{\beta_{a}\in \Delta_+}(\beta_a(\hat{\omega}))^{-2\langle w,\beta_a\rangle/\langle \beta_a,\beta_a\rangle} =\frac{1}{A_w \cdot B_w}.\nonumber
\end{equation}
\end{Definition}
We check this conjecture as far as we can and obtain  results are consistent with the results computed by  Mikhaylov's conjecture.
This conjecture is consistent with the  Conjecture \ref{c2}, since all the strings are two terms long in the weight spaces for the minuscule representations  of $ADE$ groups.

\subsection{None simple lattice Lie algebras}\label{EL}
For the fundamental representations of none simple lattice Lie algebras except the minuscule representations,  only part of the weights   satisfy  the formula (\ref{cc}).  However, there is no simple rule to fix them.

 In this subsection,  we collect   explicit formulas of solutions  for   $G_2$ group \footnote{The solutions of \textbf{KW} equation for $G_2$ group are determined in  \cite{vm}.} and $F_4$ group.   For theses  weight $w$ satisfy  the formula (\ref{cc}), we give only the factor $F_w$ which are enough  to construct the factor $W_w$. Otherwise  we give the factors  $F_w$ and $W_w$ together.

There is a   triple line in the  Dynkin diagram of $G_2$.  For the  weight $w$ does not satisfy the formula (\ref{cc}), if $w \neq [0,0,\cdots,0] $,  we find
$$W_w\cdot F_w=\langle \upsilon_{w}(\hat{\omega})|\upsilon_{w}(\hat{\omega})\rangle
\prod\limits_{\beta_{a}\in \Delta_+}(\beta_a(\hat{\omega}))^{-2\langle w,\beta_a\rangle/\langle \beta_a,\beta_a\rangle}\propto\frac{\mathfrak{n}_\mathfrak{g}}{A_w \cdot B_w}$$
where $A_w$ and  $B_w$ are defined in Eq.(\ref{ab}).  And $\mathfrak{n}_\mathfrak{g}$ is the ratio of the length squared of the long and short roots of $G$; it equals  2 for $F_4$ and  3 for $G_2$.

\begin{figure}
  \begin{center}
    \includegraphics[width=6in]{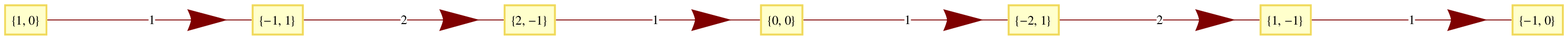}
  \end{center}
  \caption{Weights in the fundament representation $\rho_1$ of  $G_2$.   Seven  vectors $|\upsilon_{w}(\hat{\omega})\rangle$  need to be considered to compute $e^{-\chi_1}$.}
  \label{G2-10}
\end{figure}

  The weights in the first fundamental representation $\rho_{[1,0]}$ are shown in Fig.(\ref{G2-10}).  The factors  $F_w$ and $W_w$ corresponding to $w$ are given as follows
\begin{eqnarray*}
 F_{[1,0]}&=&{m_2 \left(m_1+m_2\right) \left(m_1+2 m_2\right){}^2 \left(m_1+3 m_2\right) \left(2 m_1+3 m_2\right)}\\
 F_{[-1,1]}&=&\frac{m_2}{m_1 \left(m_1+m_2\right){}^2 \left(m_1+2 m_2\right) \left(2 m_1+3 m_2\right)} \\
 F_{[2,-1]}&=&\frac{m_1 \left(m_1+m_2\right)}{m_2^2 \left(m_1^2+5 m_1 m_2+6 m_2^2\right)} \\
 F_{[-2,1]}&=&\frac{m_2^2 \left(m_1+2 m_2\right) \left(m_1+3 m_2\right)}{m_1 \left(m_1+m_2\right)} \\
 F_{[1,-1]}&=&\frac{m_1 \left(m_1+m_2\right){}^2 \left(m_1+2 m_2\right) \left(2 m_1+3 m_2\right)}{m_2}\\
 F_{[-1,0]}&=&m_2 \left(m_1+m_2\right) \left(m_1+2 m_2\right){}^2 \left(m_1+3 m_2\right) \left(2 m_1+3 m_2\right).
\end{eqnarray*}
 We find only the weight $[0,0]$ does not satisfy the  formula (\ref{cc}).
\begin{eqnarray*}
 F_{[0,0]}&=&1 \\
W_{[0,0]}&=&\frac{2}{m_2^2 \left(m_1^2+3 m_1 m_2+2 m_2^2\right){}^2}.
\end{eqnarray*}

The weights in the first fundamental representation $\rho_{[1,0]}$ are shown in Fig.(\ref{G2-01}). The factors  $F_w$ and $W_w$ corresponding to $w$ are given as follows
\begin{eqnarray*}
 F_{[0,1]}&=&{m_1 \left(m_1+m_2\right){}^3 \left(m_1+2 m_2\right){}^3 \left(m_1+3 m_2\right) \left(2 m_1+3 m_2\right){}^2}\\
F_{[3,-1]}&=&\frac{m_1}{m_2^3 \left(m_1+2 m_2\right){}^3 \left(m_1+3 m_2\right){}^2 \left(2 m_1+3 m_2\right)} \\
F_{[-3,2]}&=&\frac{m_2^3 \left(m_1+3 m_2\right)}{m_1^2 \left(m_1+m_2\right){}^3 \left(2 m_1+3 m_2\right)} \\
F_{[3,-2]}&=&\frac{m_1^2 \left(m_1+m_2\right){}^3 \left(2 m_1+3 m_2\right)}{m_2^3 \left(m_1+3 m_2\right)} \\
 F_{[-3,1]}&=&\frac{m_2^3 \left(m_1+2 m_2\right){}^3 \left(m_1+3 m_2\right){}^2 \left(2 m_1+3 m_2\right)}{m_1} \\
F_{[0,-1]}&=&m_1 \left(m_1+m_2\right){}^3 \left(m_1+2 m_2\right){}^3 \left(m_1+3 m_2\right) \left(2 m_1+3 m_2\right){}^2
\end{eqnarray*}
and
\begin{eqnarray*}
F_{[0,0]}&=&1 \\
  W_{[0,0]}&=&\frac{24 \left(m_1^2+3 m_1 m_2+3 m_2^2\right)}{m_1^2 m_2^2 \left(m_1+m_2\right){}^2 \left(2 m_1+3 m_2\right){}^2 \left(m_1^2+5 m_1 m_2+6 m_2^2\right){}^2}
\end{eqnarray*}
\begin{eqnarray*}
  F_{[1,0]}&=&\frac{1}{m_2 \left(m_1+m_2\right) \left(m_1+2 m_2\right){}^2 \left(m_1+3 m_2\right) \left(2 m_1+3 m_2\right)} \\
W_{[1,0]}&=&\frac{3 }{m_2^3 \left(m_1+m_2\right){}^3 \left(m_1+2 m_2\right){}^2 \left(2 m_1^2+9 m_1 m_2+9 m_2^2\right)}\\
 F_{[-1,1]}&=&\frac{m_2}{m_1 \left(m_1+m_2\right){}^2 \left(m_1+2 m_2\right) \left(2 m_1+3 m_2\right)} \\
 W_{[-1,1]}&=&\frac{3 }{m_1 m_2^3 \left(m_1+m_2\right){}^2 \left(m_1+2 m_2\right){}^3 \left(2 m_1+3 m_2\right)}\\
\end{eqnarray*}
\begin{eqnarray*}
 F_{[2,-1]}&=&\frac{m_1 \left(m_1+m_2\right)}{m_2^2 \left(m_1^2+5 m_1 m_2+6 m_2^2\right)} \\
 W_{[2,-1]}&=&\frac{3 }{m_1 m_2^2 \left(m_1+m_2\right){}^3 \left(m_1+2 m_2\right){}^3 \left(m_1+3 m_2\right)}\\
F_{[-2,1]}&=&\frac{m_2^2 \left(m_1+2 m_2\right) \left(m_1+3 m_2\right)}{m_1 \left(m_1+m_2\right)}\\
W_{[-2,1]}&=&\frac{3 }{m_1 m_2^2 \left(m_1+m_2\right){}^3 \left(m_1+2 m_2\right){}^3 \left(m_1+3 m_2\right)} \\
 F_{[1,-1]}&=&\frac{m_1 \left(m_1+m_2\right){}^2 \left(m_1+2 m_2\right) \left(2 m_1+3 m_2\right)}{m_2} \\
W_{[1,-1]}&=&\frac{3 }{m_1 m_2^3 \left(m_1+m_2\right){}^2 \left(m_1+2 m_2\right){}^3 \left(2 m_1+3 m_2\right)} \\
  F_{[-1,0]}&=&m_2 \left(m_1+m_2\right) \left(m_1+2 m_2\right){}^2 \left(m_1+3 m_2\right) \left(2 m_1+3 m_2\right) \\
 W_{[-1,0]}&=&\frac{3 }{m_2^3 \left(m_1+m_2\right){}^3 \left(m_1+2 m_2\right){}^2 \left(m_1+3 m_2\right) \left(2 m_1+3 m_2\right)}\\
\end{eqnarray*}
Besides the  weight $[0,0]$, the weights $[1,0]$,$[-1,1]$,$[2,-1]$, $[-2,1]$, $[1,-1]$, and  $[-1,0]$ also does not satisfy the  formula (\ref{cc}).
\begin{eqnarray*}
F_{[1,0]}\cdot W_{[1,0]} &=&\frac{3 }{m_2^2 \left(m_1+m_2\right){}^2} \\
 F_{[-1,1]}\cdot  W_{[-1,1]}&=&\frac{3}{m_2^2 \left(m_1+2 m_2\right){}^2} \\
 F_{[2,-1]}\cdot W_{[2,-1]} &=&\frac{3 }{\left(m_1+m_2\right){}^2 \left(m_1+2 m_2\right){}^2}\\
F_{[-2,1]}\cdot W_{[-2,1]} &=&\frac{3 }{\left(m_1+m_2\right){}^2 \left(m_1+2 m_2\right){}^2}\\
 F_{[1,-1]}\cdot W_{[1,-1]}&=&\frac{3 }{m_2^2 \left(m_1+2 m_2\right){}^2} \\
  F_{[-1,0]}\cdot W_{[-1,0]} &=&\frac{3 }{m_2^2 \left(m_1+m_2\right){}^2}
\end{eqnarray*}

For $F_4$ group,   we only present results of the first fundamental representation $\rho_{[1,0,0,0]}$ in order to save space.
\begin{tiny}
\begin{eqnarray*}
&&\textrm{exp}(-\chi_1)\\
&=&2^{-B_{[1,0,0,0]}}(EWF^{'}(\sigma)+WF_{[0,0,0,0]})\\
&=&\frac{1}{512}(\frac{3 \cosh{(-2 \sigma  m_2)}}{m_1 m_2^2 \left(m_1+m_2\right){}^3 \left(m_1+2 m_2\right){}^3 \left(m_1+3 m_2\right)}+\frac{\cosh{(2 \sigma  \left(2 m_1+3 m_2\right)}}{m_1 \left(m_1+m_2\right){}^3 \left(m_1+2 m_2\right){}^3 \left(m_1+3 m_2\right) \left(2 m_1+3 m_2\right){}^2}-\\
&&\frac{3 \cosh{(2 \sigma  \left(m_1+m_2\right))}}{m_1 m_2^3 \left(m_1+m_2\right){}^2 \left(m_1+2 m_2\right){}^3 \left(2 m_1+3 m_2\right)}-\frac{\cosh{(2 \sigma  \left(m_1+3 m_2\right))}}{m_1 m_2^3 \left(m_1+2 m_2\right){}^3 \left(m_1+3 m_2\right){}^2 \left(2 m_1+3 m_2\right)}+\\
&&\frac{\cosh{(2 \sigma  m_1)}}{m_1^2 m_2^3 \left(m_1+m_2\right){}^3 \left(m_1+3 m_2\right) \left(2 m_1+3 m_2\right)}+\frac{3 \cosh{(2 \sigma  \left(m_1+2 m_2\right))}}{m_2^3 \left(m_1+m_2\right){}^3 \left(m_1+2 m_2\right){}^2 \left(m_1+3 m_2\right) \left(2 m_1+3 m_2\right)}-\\
&&\frac{12 \left(m_1^2+3 m_1 m_2+3 m_2^2\right)}{m_1^2 m_2^2 \left(m_1+m_2\right){}^2 \left(m_1+2 m_2\right){}^2 \left(m_1+3 m_2\right){}^2 \left(2 m_1+3 m_2\right){}^2}).
\end{eqnarray*}
\end{tiny}
Only the weight $[0,0,0,0]$ does not satisfy the  formula (\ref{cc}). By  using formula (\ref{cs}),  it is easy to find
\begin{tiny}
\begin{eqnarray*}
WF_{[0,0,0,0]}=-\frac{12 \left(m_1^2+3 m_1 m_2+3 m_2^2\right)}{m_1^2 m_2^2 \left(m_1+m_2\right){}^2 \left(m_1+2 m_2\right){}^2 \left(m_1+3 m_2\right){}^2 \left(2 m_1+3 m_2\right){}^2}).
\end{eqnarray*}
\end{tiny}

\section{Summary and open problems}
In \cite{vm}, Mikhaylov conjectured the solutions of \textbf{KW} equations for  a boundary  't Hooft operator. In order to prove this conjecture, one   need  to check the boundary condition (\ref{bbc}).
However, there are two computational difficulties to construct the solutions Eq.(\ref{kwe}) explicitly. One difficulty  relate to the commutation  of  generators of Lie algebra  in $W_w$.  With the rank of Lie algebra  $\mathfrak{g}$ increasing,     the  commutation work of the operators   increases rapidly.  We  derived  an identity ({Proposition \ref{pro}})  which  simplifies  the calculation effectively. The computational efficiency is improved remarkably, since the factor $(\lambda_i-(n-1)-\sum_{b=1}^{m}A_{j_b,i})$ vanish  from time to time in the computation process. The other difficulty involves the number of paths $\textrm{\textbf{s}}$  reaching a weight in the fundamental representation from the highest weight.  With the rank of $\mathfrak{g}$ increasing, the number of paths $\textrm{\textbf{s}}$ in the fundament representation   as well as the number of  weight $w$   increase rapidly. For the weights in the minuscule representations and certain weights in the fundamental representations for gauge groups of $ADE$ type,  we conjecture  a formula to  rewrite the  factors $W_w\cdot F_w$ by the  co-prime  numerator $A_w$ and denominator $B_w$ of $F_w$, thus bypassing the  above two computational  difficulties coded in the factor $W_w$.

We have to point out that not all the weights of a fundamental representation are in a string of two elements along a simple root except the minuscule fundamental representations. Thus we can only simplify the constructions of the factor  $E_w\cdot W_w\cdot F_w$ for parts of weights for most fundamental representations. Notwithstanding its limitation, the Conjecture \ref{c2} and Conjecture \ref{mmd} are helpful in some special case.  According to the table of minuscule fundamental weights in Section \ref{solution1}, weights in all the fundament representations of $A_n$ algebra satisfy Eq.(\ref{ab}).  So the solutions of  \textbf{KW} equations  for $A_n$ algebra can be constructed completely  using the identity (\ref{ab}).


Clearly more work is needed.  The proof of the formula  of  solutions (\ref{kwe}) for  general gauge group $G$ is still an open problem. The conjecture \ref{c2} also need to be proved. The conjecture \ref{mmd}   may be proved   by following   Mikhaylov's proof in the $A_n$ case. It is also interesting  to construct solutions of \textbf{KW} equations for the boundary surface operator of arbitrary gauge group $G$ on a half space.  Instead of one side boundary, we can consider a two-sided problem  on $\mathbb{R}^3\times I$, where $I$ is a compact interval with 't Hooft operator or surface operator in the boundaries \cite{5knots}. We can also consider the case when  $\mathbb{R}^3$  is replaced by $S^3$.

\section*{Acknowledgments}
We would like to thank Xian Gao  and Song He for  many helpful discussions.  They also relied heavily on the softwares LiE \footnote{It can be downloaded from    http://www-math.univ-poitiers.fr/~maavl/LiE/.} and Mathematic.   This work was supported by a grant from  the Postdoctoral Foundation of Zhejiang Province.

\appendix
\section{Summary of some relevant results for $ADE$ groups}\label{B}
In the appendix of \cite{vm}, Mikhaylov collected solutions of \textbf{KW} equations for the algebras $A_1$, $A_2$, $A_3$, $B_2$ and $G_2$. In this appendix,  we collect  more explicit formulas for the 't~Hooft operator solutions for other algebras.  We present the completely solutions for $A_4$ and $D_4$. We  check these solutions   by the conjecture  \ref{c2},  getting  completely consistency results.
 The solutions for   $E_6, E_7$, and $E_8$ are not presented here since even the simplest  factor $e^{-\chi_1(\sigma)}$ for $E_6$  need more then five pages to record.

$A_4,$ $[1,0,0,0]$
\begin{tiny}
\begin{eqnarray*}
&&\textrm{exp}(-\chi_1)=\\
&&\frac{1}{64}(-\frac{e^{-\frac{2}{5} \sigma  \left(m_1+2 m_2+3 m_3-m_4\right)}}{m_3 \left(m_2+m_3\right) \left(m_1+m_2+m_3\right) m_4}+\frac{e^{-\frac{2}{5} \sigma  \left(m_1+2 m_2-2 m_3-m_4\right)}}{m_2 \left(m_1+m_2\right) m_3 \left(m_3+m_4\right)}-\frac{e^{-\frac{2}{5} \sigma  \left(m_1-3 m_2-2 m_3-m_4\right)}}{m_1 m_2 \left(m_2+m_3\right) \left(m_2+m_3+m_4\right)}+\\
&&\frac{e^{\frac{2}{5} \sigma  \left({4 m_1}+{3 m_2}+{2 m_3}+{m_4}\right)}}{m_1 \left(m_1+m_2\right) \left(m_1+m_2+m_3\right) \left(m_1+m_2+m_3+m_4\right)}+
\frac{e^{-\frac{2}{5} \sigma  \left(m_1+2 m_2+3 m_3+4 m_4\right)}}{m_4 \left(m_3+m_4\right) \left(m_2+m_3+m_4\right) \left(m_1+m_2+m_3+m_4\right)})
\end{eqnarray*}
\end{tiny}

$[0, 1, 0, 0]:$
\begin{tiny}
\begin{eqnarray*}
&&\textrm{exp}(-\chi_2)=\\
&&\frac{1}{64}(\frac{e^{-\frac{2}{5} \sigma  \left(2 m_1+4 m_2+m_3-2 m_4\right)}}{m_2 \left(m_1+m_2\right) \left(m_2+m_3\right) \left(m_1+m_2+m_3\right) m_4 \left(m_3+m_4\right)}-\frac{e^{-\frac{2}{5} \sigma  \left(2 m_1-m_2+m_3-2 m_4\right)}}{m_1 m_2 m_3 \left(m_1+m_2+m_3\right) m_4 \left(m_2+m_3+m_4\right)}+\\
&&\frac{e^{-\frac{2}{5} \sigma  \left(2 m_1-m_2-2 \left(2 m_3+m_4\right)\right)}}{m_1 \left(m_1+m_2\right) m_3 \left(m_2+m_3\right) \left(m_3+m_4\right) \left(m_2+m_3+m_4\right)}-
\frac{e^{\frac{2}{5} \sigma  \left(3 m_1+m_2-m_3-3 m_4\right)}}{m_1 \left(m_1+m_2\right) \left(m_1+m_2+m_3\right) m_4 \left(m_3+m_4\right) \left(m_2+m_3+m_4\right)}+\\
&&\frac{e^{\frac{2}{5} \sigma  \left(3 m_1+m_2-m_3+2 m_4\right)}}{m_1 \left(m_1+m_2\right) m_3 \left(m_2+m_3\right) m_4 \left(m_1+m_2+m_3+m_4\right)}-\frac{e^{\frac{2}{5} \sigma  \left(3 m_1+m_2+4 m_3+2 m_4\right)}}{m_1 m_2 m_3 \left(m_1+m_2+m_3\right) \left(m_3+m_4\right) \left(m_1+m_2+m_3+m_4\right)}+\\
&&\frac{e^{-\frac{2}{5} \sigma  \left(2 m_1-m_2+m_3+3 m_4\right)}}{m_1 m_2 \left(m_2+m_3\right) m_4 \left(m_3+m_4\right) \left(m_1+m_2+m_3+m_4\right)}\\
&&+\frac{e^{\frac{2}{5} \sigma  \left({3 m_1}+{6 m_2}+{4 m_3}+{2 m_4}\right)}}{m_2 \left(m_1+m_2\right) \left(m_2+m_3\right) \left(m_1+m_2+m_3\right) \left(m_2+m_3+m_4\right) \left(m_1+m_2+m_3+m_4\right)}-\\
&&\frac{e^{-\frac{2}{5} \sigma  \left(2 m_1+4 m_2+m_3+3 m_4\right)}}{m_2 \left(m_1+m_2\right) m_3 m_4 \left(m_2+m_3+m_4\right) \left(m_1+m_2+m_3+m_4\right)}\\
&&+\frac{e^{-\frac{2}{5} \sigma  \left(2 m_1+4 m_2+6 m_3+3 m_4\right)}}{m_3 \left(m_2+m_3\right) \left(m_1+m_2+m_3\right) \left(m_3+m_4\right) \left(m_2+m_3+m_4\right) \left(m_1+m_2+m_3+m_4\right)})
\end{eqnarray*}
\end{tiny}

$[0, 0, 1, 0]: $
\begin{tiny}
\begin{eqnarray*}
&&\textrm{exp}(-\chi_3)=\\
&&\frac{1}{64}(\frac{e^{\frac{2}{5} \sigma  \left(2 m_1+4 m_2+m_3-2 m_4\right)}}{m_2 \left(m_1+m_2\right) \left(m_2+m_3\right) \left(m_1+m_2+m_3\right) m_4 \left(m_3+m_4\right)}-\frac{e^{\frac{2}{5} \sigma  \left(2 m_1-m_2+m_3-2 m_4\right)}}{m_1 m_2 m_3 \left(m_1+m_2+m_3\right) m_4 \left(m_2+m_3+m_4\right)}+\\
&&\frac{e^{\frac{2}{5} \sigma  \left(2 m_1-m_2-2 \left(2 m_3+m_4\right)\right)}}{m_1 \left(m_1+m_2\right) m_3 \left(m_2+m_3\right) \left(m_3+m_4\right) \left(m_2+m_3+m_4\right)}-\frac{e^{-\frac{2}{5} \sigma  \left(3 m_1+m_2-m_3-3 m_4\right)}}{m_1 \left(m_1+m_2\right) \left(m_1+m_2+m_3\right) m_4 \left(m_3+m_4\right) \left(m_2+m_3+m_4\right)}\\
&&+\frac{e^{-\frac{2}{5} \sigma  \left(3 m_1+m_2-m_3+2 m_4\right)}}{m_1 \left(m_1+m_2\right) m_3 \left(m_2+m_3\right) m_4 \left(m_1+m_2+m_3+m_4\right)}-\frac{e^{-\frac{2}{5} \sigma  \left(3 m_1+m_2+4 m_3+2 m_4\right)}}{m_1 m_2 m_3 \left(m_1+m_2+m_3\right) \left(m_3+m_4\right) \left(m_1+m_2+m_3+m_4\right)}+\\
&&\frac{e^{\frac{2}{5} \sigma  \left(2 m_1-m_2+m_3+3 m_4\right)}}{m_1 m_2 \left(m_2+m_3\right) m_4 \left(m_3+m_4\right) \left(m_1+m_2+m_3+m_4\right)}\\
&&+\frac{e^{-\frac{2}{5} \sigma  \left(3 m_1+6 m_2+4 m_3+2 m_4\right)}}{m_2 \left(m_1+m_2\right) \left(m_2+m_3\right) \left(m_1+m_2+m_3\right) \left(m_2+m_3+m_4\right) \left(m_1+m_2+m_3+m_4\right)}-\\
&&\frac{e^{\frac{2}{5} \sigma  \left(2 m_1+4 m_2+m_3+3 m_4\right)}}{m_2 \left(m_1+m_2\right) m_3 m_4 \left(m_2+m_3+m_4\right) \left(m_1+m_2+m_3+m_4\right)}\\
&&+\frac{e^{\frac{2 }{5}\sigma  \left({2 m_1}+{4 m_2}+{6 m_3}+{3 m_4}\right)}}{m_3 \left(m_2+m_3\right) \left(m_1+m_2+m_3\right) \left(m_3+m_4\right) \left(m_2+m_3+m_4\right) \left(m_1+m_2+m_3+m_4\right)})
\end{eqnarray*}
\end{tiny}

$[0,0,0,1]:$
\begin{tiny}
\begin{eqnarray*}
&&\textrm{exp}(-\chi_4)=\\
&&\frac{1}{64}(-\frac{e^{\frac{2}{5} \sigma  \left(m_1+2 m_2+3 m_3-m_4\right)}}{m_3 \left(m_2+m_3\right) \left(m_1+m_2+m_3\right) m_4}+\frac{e^{\frac{2}{5} \sigma  \left(m_1+2 m_2-2 m_3-m_4\right)}}{m_2 \left(m_1+m_2\right) m_3 \left(m_3+m_4\right)}-\frac{e^{\frac{2}{5} \sigma  \left(m_1-3 m_2-2 m_3-m_4\right)}}{m_1 m_2 \left(m_2+m_3\right) \left(m_2+m_3+m_4\right)}+\\
&&\frac{e^{-\frac{2}{5} \sigma  \left(4 m_1+3 m_2+2 m_3+m_4\right)}}{m_1 \left(m_1+m_2\right) \left(m_1+m_2+m_3\right) \left(m_1+m_2+m_3+m_4\right)}+\frac{e^{\frac{2}{5}\sigma  \left({m_1}+{2 m_2}+{3 m_3}+{4 m_4}\right)}}{m_4 \left(m_3+m_4\right) \left(m_2+m_3+m_4\right) \left(m_1+m_2+m_3+m_4\right)})
\end{eqnarray*}
\end{tiny}

For ${D_4}$, the cartan matrix  is
$$
\begin{pmatrix}
 2 & -1 & 0 & 0 \\
 -1 & 2 & -1 & -1 \\
 0 & -1 & 2 & 0 \\
 0 & -1 & 0 & 2\\
\end{pmatrix}.
$$

 The weights in the fundamental representation $\rho_{[0,1,0,0]}$ are shown in Fig.(\ref{D4-0100}).

 $[1,0,0,0]$:
\begin{tiny}
\begin{eqnarray*}
&&\textrm{exp}(-\chi_1)=\\
&&-\frac{\cosh{( \sigma  \left({m_3}-{m_4}\right))}}{m_3 \left(m_2+m_3\right) \left(m_1+m_2+m_3\right) m_4 \left(m_2+m_4\right) \left(m_1+m_2+m_4\right)}\\
&&+\frac{\cosh{(\sigma  \left({m_3}+{m_4}\right))}}{m_2 \left(m_1+m_2\right) m_3 m_4 \left(m_2+m_3+m_4\right) \left(m_1+m_2+m_3+m_4\right)}\\
&&-\frac{\cosh{(\sigma  \left(2m_2+{m_3}+{m_4}\right))}}{m_1 m_2 \left(m_2+m_3\right) \left(m_2+m_4\right) \left(m_2+m_3+m_4\right) \left(m_1+2 m_2+m_3+m_4\right)}\\
&&+\frac{\cosh{( \sigma  \left(2m_1+2m_2+{m_3}+{m_4}\right))}}{m_1 \left(m_1+m_2\right) \left(m_1+m_2+m_3\right) \left(m_1+m_2+m_4\right) \left(m_1+m_2+m_3+m_4\right) \left(m_1+2 m_2+m_3+m_4\right)}
\end{eqnarray*}
\end{tiny}

$ [0,1,0,0]$:
\begin{tiny}
\begin{eqnarray*}
&&\textrm{exp}(-\chi_2)\\
&=&2^{-B_{[0,1,0,0]}}(EWF^{'}(\sigma)+WF_{[0,0,0,0]})\\
&=&\frac{1}{32}(\frac{\cosh{(2 \sigma  m_1)}}{m_1^2 m_2 \left(m_1+m_2\right) \left(m_2+m_3\right) \left(m_1+m_2+m_3\right) \left(m_2+m_4\right) \left(m_1+m_2+m_4\right) \left(m_2+m_3+m_4\right) \left(m_1+m_2+m_3+m_4\right)}\\
&&+\frac{\cosh{(-2 \sigma  m_3)}}{m_2 \left(m_1+m_2\right) m_3^2 \left(m_2+m_3\right) \left(m_1+m_2+m_3\right) \left(m_2+m_4\right) \left(m_1+m_2+m_4\right) \left(m_2+m_3+m_4\right) \left(m_1+m_2+m_3+m_4\right)}+\\
&&\frac{\cosh{(-2 \sigma  m_4)}}{m_2 \left(m_1+m_2\right) \left(m_2+m_3\right) \left(m_1+m_2+m_3\right) m_4^2 \left(m_2+m_4\right) \left(m_1+m_2+m_4\right) \left(m_2+m_3+m_4\right) \left(m_1+m_2+m_3+m_4\right)}+\\
&&\frac{\cosh{(2 \sigma  \left(-m_1-2 m_2-m_3-m_4\right))}}{m_2 \left(m_1+m_2\right) \left(m_2+m_3\right) \left(m_1+m_2+m_3\right) \left(m_2+m_4\right) \left(m_1+m_2+m_4\right) }\cdot\\
&&\frac{1}{\left(m_2+m_3+m_4\right) \left(m_1+m_2+m_3+m_4\right) \left(m_1+2 m_2+m_3+m_4\right){}^2}-\\
&&\frac{\cosh{(2 \sigma  \left(-m_1-m_2\right))}}{m_1 m_2 \left(m_1+m_2\right){}^2 m_3 \left(m_1+m_2+m_3\right) m_4 \left(m_1+m_2+m_4\right) \left(m_2+m_3+m_4\right) \left(m_1+2 m_2+m_3+m_4\right)}-\\
&&\frac{\cosh{(2 \sigma  \left(-m_2-m_3\right))}}{m_1 m_2 m_3 \left(m_2+m_3\right){}^2 \left(m_1+m_2+m_3\right) m_4 \left(m_1+m_2+m_4\right) \left(m_2+m_3+m_4\right) \left(m_1+2 m_2+m_3+m_4\right)}-\\
&&\frac{\cosh{(2 \sigma  \left(-m_2-m_4\right))}}{m_1 m_2 m_3 \left(m_1+m_2+m_3\right) m_4 \left(m_2+m_4\right){}^2 \left(m_1+m_2+m_4\right) \left(m_2+m_3+m_4\right) \left(m_1+2 m_2+m_3+m_4\right)}-\\
&&\frac{\cosh{(2 \sigma  \left(-m_1-m_2-m_3-m_4\right))}}{m_1 m_2 m_3 \left(m_1+m_2+m_3\right) m_4 \left(m_1+m_2+m_4\right) \left(m_2+m_3+m_4\right) \left(m_1+m_2+m_3+m_4\right){}^2 \left(m_1+2 m_2+m_3+m_4\right)}-\\
&&\frac{\cosh{(2 \sigma  m_2)}}{m_1 m_2^2 \left(m_1+m_2\right) m_3 \left(m_2+m_3\right) m_4 \left(m_2+m_4\right) \left(m_1+m_2+m_3+m_4\right) \left(m_1+2 m_2+m_3+m_4\right)}+\\
&&\frac{\cosh{(2 \sigma  \left(-m_1-m_2-m_3\right))}}{m_1 \left(m_1+m_2\right) m_3 \left(m_2+m_3\right) \left(m_1+m_2+m_3\right){}^2 m_4 \left(m_2+m_4\right) \left(m_1+m_2+m_3+m_4\right) \left(m_1+2 m_2+m_3+m_4\right)}+\\
&&\frac{\cosh{(2 \sigma  \left(-m_1-m_2-m_4\right))}}{m_1 \left(m_1+m_2\right) m_3 \left(m_2+m_3\right) m_4 \left(m_2+m_4\right) \left(m_1+m_2+m_4\right){}^2 \left(m_1+m_2+m_3+m_4\right) \left(m_1+2 m_2+m_3+m_4\right)}+\\
&&\frac{\cosh{(2 \sigma  \left(-m_2-m_3-m_4\right))}}{m_1 \left(m_1+m_2\right) m_3 \left(m_2+m_3\right) m_4 \left(m_2+m_4\right) \left(m_2+m_3+m_4\right){}^2 \left(m_1+m_2+m_3+m_4\right) \left(m_1+2 m_2+m_3+m_4\right)}\\
&&-WF_{[0,0,0,0]})
\end{eqnarray*}
\end{tiny}
We split the term $\sum\limits_{w \in \Delta_s}\left[E_w\cdot W_w\cdot F_w\right]$  in (\ref{nkwe}) into two parts.  The term $E_w\cdot W_w\cdot F_w$ with the weight $[0,\cdots,0]$ is denoted as $WF_{[0,0,0,0]}$.  And the other  terms are denoted as $EWF^{'}(\sigma)$.

$ [0,0,1,0]$:
\begin{tiny}
\begin{eqnarray*}
&&\textrm{exp}(-\chi_3)=\\
&&\frac{1}{1024}(-\frac{e^{\sigma  \left(m_1-m_4\right)}}{m_1 \left(m_1+m_2\right) \left(m_1+m_2+m_3\right) m_4 \left(m_2+m_4\right) \left(m_2+m_3+m_4\right)}-\\
&&\frac{e^{\sigma  \left(-m_1+m_4\right)}}{m_1 \left(m_1+m_2\right) \left(m_1+m_2+m_3\right) m_4 \left(m_2+m_4\right) \left(m_2+m_3+m_4\right)}+\frac{e^{-\sigma  \left(m_1+m_4\right)}}{m_1 m_2 \left(m_2+m_3\right) m_4 \left(m_1+m_2+m_4\right) \left(m_1+m_2+m_3+m_4\right)}+\\
&&\frac{e^{\sigma  \left(m_1+m_4\right)}}{m_1 m_2 \left(m_2+m_3\right) m_4 \left(m_1+m_2+m_4\right) \left(m_1+m_2+m_3+m_4\right)}\\
&&-\frac{e^{-\sigma  \left(m_1+2 m_2+m_4\right)}}{m_2 \left(m_1+m_2\right) m_3 \left(m_2+m_4\right) \left(m_1+m_2+m_4\right) \left(m_1+2 m_2+m_3+m_4\right)}-\\
&&\frac{e^{\sigma  \left(m_1+2 m_2+m_4\right)}}{m_2 \left(m_1+m_2\right) m_3 \left(m_2+m_4\right) \left(m_1+m_2+m_4\right) \left(m_1+2 m_2+m_3+m_4\right)}+\\
&&\frac{e^{-\sigma  \left(m_1+2 m_2+2 m_3+m_4\right)}}{m_3 \left(m_2+m_3\right) \left(m_1+m_2+m_3\right) \left(m_2+m_3+m_4\right) \left(m_1+m_2+m_3+m_4\right) \left(m_1+2 m_2+m_3+m_4\right)}+\\
&&\frac{e^{\sigma  \left(m_1+2 m_2+2 m_3+m_4\right)}}{m_3 \left(m_2+m_3\right) \left(m_1+m_2+m_3\right) \left(m_2+m_3+m_4\right) \left(m_1+m_2+m_3+m_4\right) \left(m_1+2 m_2+m_3+m_4\right)})
\end{eqnarray*}
\end{tiny}

$ [0,0,0,1]$:
\begin{tiny}
\begin{eqnarray*}
&&\textrm{exp}(-\chi_4)=\\
&&\frac{1}{1024}(-\frac{e^{\sigma  \left(m_1-m_3\right)}}{m_1 \left(m_1+m_2\right) m_3 \left(m_2+m_3\right) \left(m_1+m_2+m_4\right) \left(m_2+m_3+m_4\right)}\\
&&-\frac{e^{\sigma  \left(-m_1+m_3\right)}}{m_1 \left(m_1+m_2\right) m_3 \left(m_2+m_3\right) \left(m_1+m_2+m_4\right) \left(m_2+m_3+m_4\right)}+\frac{e^{-\sigma  \left(m_1+m_3\right)}}{m_1 m_2 m_3 \left(m_1+m_2+m_3\right) \left(m_2+m_4\right) \left(m_1+m_2+m_3+m_4\right)}\\
&&+\frac{e^{\sigma  \left(m_1+m_3\right)}}{m_1 m_2 m_3 \left(m_1+m_2+m_3\right) \left(m_2+m_4\right) \left(m_1+m_2+m_3+m_4\right)}\\
&&-\frac{e^{-\sigma  \left(m_1+2 m_2+m_3\right)}}{m_2 \left(m_1+m_2\right) \left(m_2+m_3\right) \left(m_1+m_2+m_3\right) m_4 \left(m_1+2 m_2+m_3+m_4\right)}-\\
&&\frac{e^{\sigma  \left(m_1+2 m_2+m_3\right)}}{m_2 \left(m_1+m_2\right) \left(m_2+m_3\right) \left(m_1+m_2+m_3\right) m_4 \left(m_1+2 m_2+m_3+m_4\right)}+\\
&&\frac{e^{-\sigma  \left(m_1+2 m_2+m_3+2 m_4\right)}}{m_4 \left(m_2+m_4\right) \left(m_1+m_2+m_4\right) \left(m_2+m_3+m_4\right) \left(m_1+m_2+m_3+m_4\right) \left(m_1+2 m_2+m_3+m_4\right)}+\\
&&\frac{e^{\sigma  \left(m_1+2 m_2+m_3+2 m_4\right)}}{m_4 \left(m_2+m_4\right) \left(m_1+m_2+m_4\right) \left(m_2+m_3+m_4\right) \left(m_1+m_2+m_3+m_4\right) \left(m_1+2 m_2+m_3+m_4\right)})
\end{eqnarray*}
\end{tiny}

\end{document}